\documentclass[lp,12pt]{elsarticle}
\usepackage{amsmath,amsfonts,multirow,graphicx,subfigure,mathtools,setspace,tabularx,rotating,pdflscape,bbm}
\biboptions{}

\begin{document}


\title{Non stationary multifractality in stock returns}


\author{Raffaello Morales}
\ead{raffaello.morales@kcl.ac.uk}
\address{Department of Mathematics, King's College London, Strand, WC2R 2LS, London, UK}
\author{T. Di Matteo}
\address{Department of Mathematics, King's College London, Strand, WC2R 2LS, London, UK}
\author{Tomaso Aste}
\address{Department of Computer Science, University College London, Gower Street, WC1E 6BT, London, UK}


\begin{abstract}
We perform an extensive empirical analysis of scaling properties of equity returns, suggesting that financial data show time varying multifractal properties. This is obtained by comparing empirical observations of the weighted generalised Hurst exponent (wGHE) with time series simulated via Multifractal Random Walk (MRW) by Bacry \textit{et al.} [\textit{E.Bacry, J.Delour and J.Muzy, Phys.Rev.E \,{\bf 64} 026103, 2001}]. While dynamical wGHE computed on synthetic MRW series is consistent with a scenario where multifractality is constant over time, fluctuations in the dynamical wGHE observed in empirical data are not in agreement with a MRW with constant intermittency parameter. We test these hypotheses of constant multifractality considering different specifications of MRW model with fatter tails: in all cases considered, although the thickness of the tails accounts for most of anomalous fluctuations of multifractality, still cannot fully explain the observed fluctuations.  
\end{abstract}


\maketitle

\section{Introduction}
The concept of multifractality in the context of finance has received much attention in the econophysics literature \cite{mantegna2000introduction} over the last two decades. Many empirical studies have investigated financial data scaling behaviour \cite{kantelhardt2002multifractal, matteo2005long, muzy2000modelling, barunik2010hurst, di2007multi,barunik2012understanding, kristoufek2011multifractal,jiang2008multifractality,jiang2008multifractal} and several models have been proposed to account for the observed multifractal features \cite{bacry2001multifractal,bouchaud2000apparent,liu2008multifractality,calvet2002multifractality,liu2007true,mandelbrot1997multifractal,lux2008markov,ding1993long,borland2005multi}. 
Multifractal behaviour has become a downright stylised fact of financial market data \cite{bouchaud2003theory,dacorogna2001introduction}, being observed across several classes of assets: from daily stock prices to foreign exchange rates and composite indices \cite{di2007multi,matteo2005long,di2003scaling,bartolozzi2007multi,zhou2009components}. 
\newline Multifractality is also particularly appealing for modelling financial markets as it offers a simple behavioural interpretation: looking at the volatility at different time scales is a very natural way to assess the impact of heterogeneous agents in the market and therefore any measure of scaling behaviour can convey information about the efficiency of a given market \cite{dacorogna2001introduction}. In this regard some authors have suggested, and confirmed through extensive empirical studies, that scaling exponents can be representative of the stage of development of a market \cite{di2003scaling,matteo2005long}. The same works have convincingly shown that, through a hierarchy of scaling exponents, it is possible to classify markets according to their degree of development with emerging markets exhibiting scaling exponents significantly larger than those observed in developed markets. 
\newline Let us recall that a process $X(t)$ with stationary increments is called multifractal if the following scaling law is observed
\begin{equation}\label{scaling1}
\mathbb{E}(|X(t+\tau)-X(t)|^{q})\sim c_{q}\tau^{\zeta_{q}}, 
\end{equation}
where $c_{q}$ is a constant and $\zeta_{q}$ is a non-linear function of $q$, called the scaling function. The departure of $\zeta_{q}$ from linearity is what distinguishes multifractal processes from uni-scaling processes and the degree of non-linearity of the scaling function is accounted for by the \textit{intermittency coefficient}, defined as $\lambda^2=\zeta''_{0}$ \cite{bacry2001multifractal}. Note that the last definition requires the scaling function to have a second derivative well defined in $q=0$ in order for the intermittency coefficient to be defined. (In practice, it is extremely difficult to estimate this coefficient from empirical measures.) For multifractal processes, the scaling in equation ($\ref{scaling1}$) holds for small $\tau$, with $\tau/T\ll1$, where $T$ is some larger scale called the \textit{integral scale} \cite{bacry2001multifractal}. This means that if the integral scale of the process is not large enough compared to the resolution of the increments, although the scaling may in principle hold, the definition doesn't hold any longer.   
\newline In order to estimate the scaling function from real data one resorts to the scaling of the empirical moments
\begin{equation}\label{scaling2}
\frac{1}{N-\tau+1}\sum_{t=0}^{N-\tau}|r_{t,\tau}|^q\sim \tau^{\zeta^{*}_{q}},
\end{equation}
where we denote $r_{t,\tau}=\log p_{t+\tau}-\log p_{t}$ the log-return at time $t$ and scale $\tau$, with $p_{t}$ the asset price at time $t$, $\zeta^{*}_{q}$ the empirical scaling function and $N$ the length of the time series. It is well known \cite{muzy2006extreme,muzy2008uncovering} that the empirically estimated $\zeta^{*}_{q}$ is significant only for small values of $q$ and one therefore needs to be careful in interpreting the scaling beyond a certain $q$. When the scaling (\ref{scaling2}) is observed, one defines the generalised Hurst exponent $H(q)$ (GHE) via 
\begin{equation}\label{zeta_Hurst}
\zeta^{*}_{q}=qH(q).
\end{equation}
The exponent $H(q)$ is non-linear in $q$ for multi-scaling processes, whereas it reduces to a constant $H$ if the process is uni-scaling. 
\newline The time evolution of the scaling function, measured via the GHE, can be useful to track time varying properties of the market. For this reason $\zeta^{*}_{q}$ has also been studied dynamically via the time dependent (or local) Hurst exponent \cite{carbone2004time,grech2008local,barunik2010hurst}. In a recent publication \cite{morales2012dynamical} the authors have observed large fluctuations in multifractality measured via the generalised Hurst exponent in empirical daily data across different stock sectors. Specifically, the authors considered as a measure of multifractal behaviour the quantity 
\begin{equation}
\Delta H^{w}(q,q^{'})= H^{w}(q)-H^{w}(q^{'}), \qquad q\neq q^{'},
\end{equation}
where $H^{w}(q)$ is the weighted generalized Hurst exponent (wGHE) \cite{morales2012dynamical}. The weighting procedure incrementally damps effects from past return fluctuations. The dynamical evolution of $\Delta H^{w}(q,q^{'})$ (which we shall label $\Delta H^{w}_{t}(q,q^{'})$) is useful to track changes in multifractality occurring over time.
\newline When facing the task of ascertaining the nature of dynamical fluctuations in these quantities, the subtle issue is being able to distinguish between spurious statistical fluctuations, which are due to the finiteness of the sample and noise, and true structural changes in the underlying multifractal process. In this paper we study the problem of validating dynamical fluctuations of the scaling functions, performing an empirical analysis of stock returns and comparing their properties with those of synthetic multifractal series. 
Among all existing models we focus on the Multifractal Random Walk (MRW) introduced by Bacry \textit{et al.} \cite{bacry2001multifractal} because of its parsimonious formulation and its success in the econophysics literature.
\newline This paper is organised as follows: in Section \ref{sec2} we review the main properties of the MRW and establish the connection with the generalised Hurst exponent. In Section \ref{sec3}, after introducing the statistical testing procedure, we report the main findings on the varying multifractality of empirical stock returns data. A summary and conclusive remarks are drawn in Section \ref{sec6}. 
\section{\label{sec2} Generalities on Multifractal random walk}
\begin{figure}
\mbox{\subfigure{\includegraphics[width= 0.5\columnwidth]{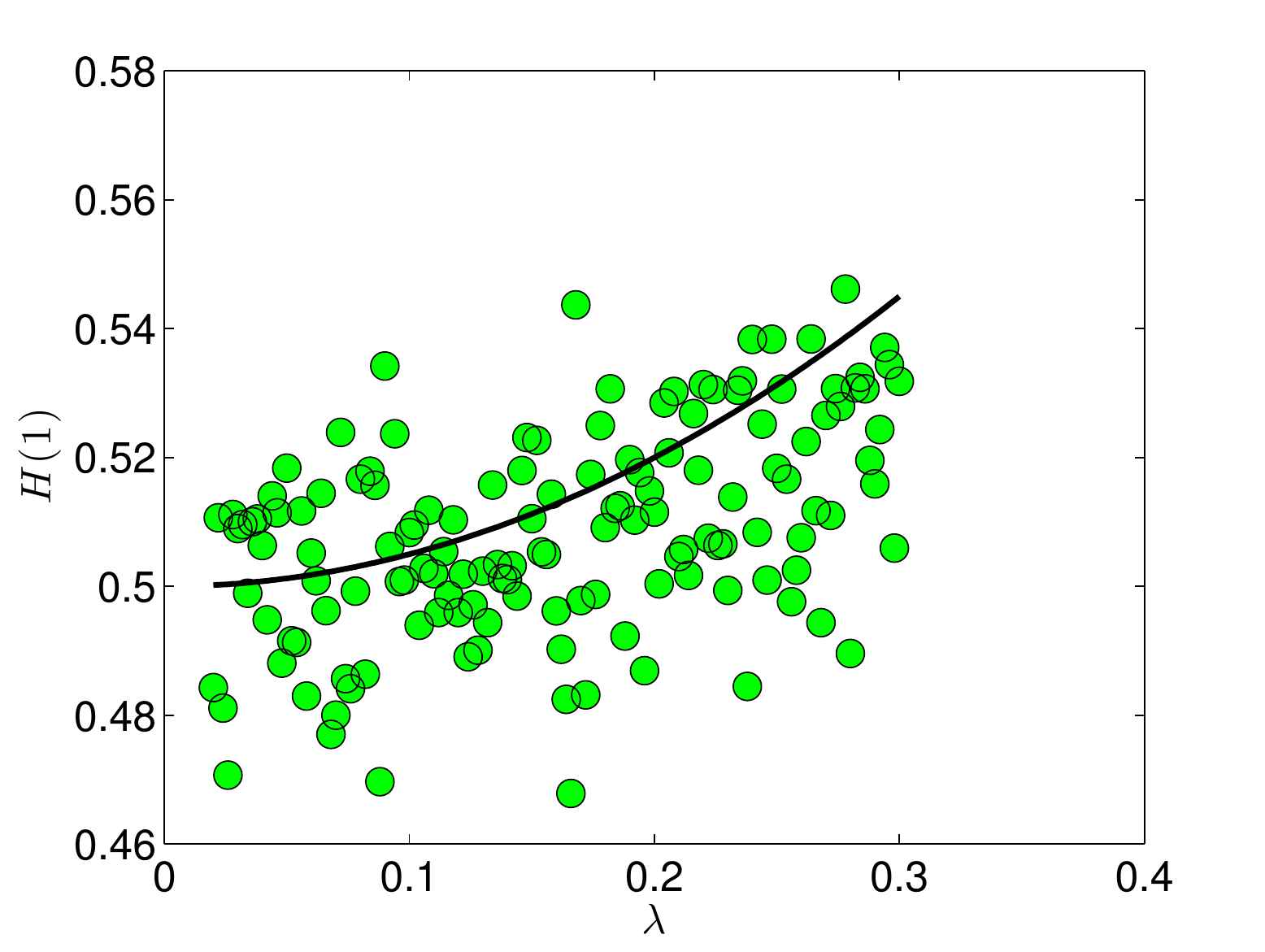}} \quad
\subfigure{\includegraphics[width=0.5\columnwidth]{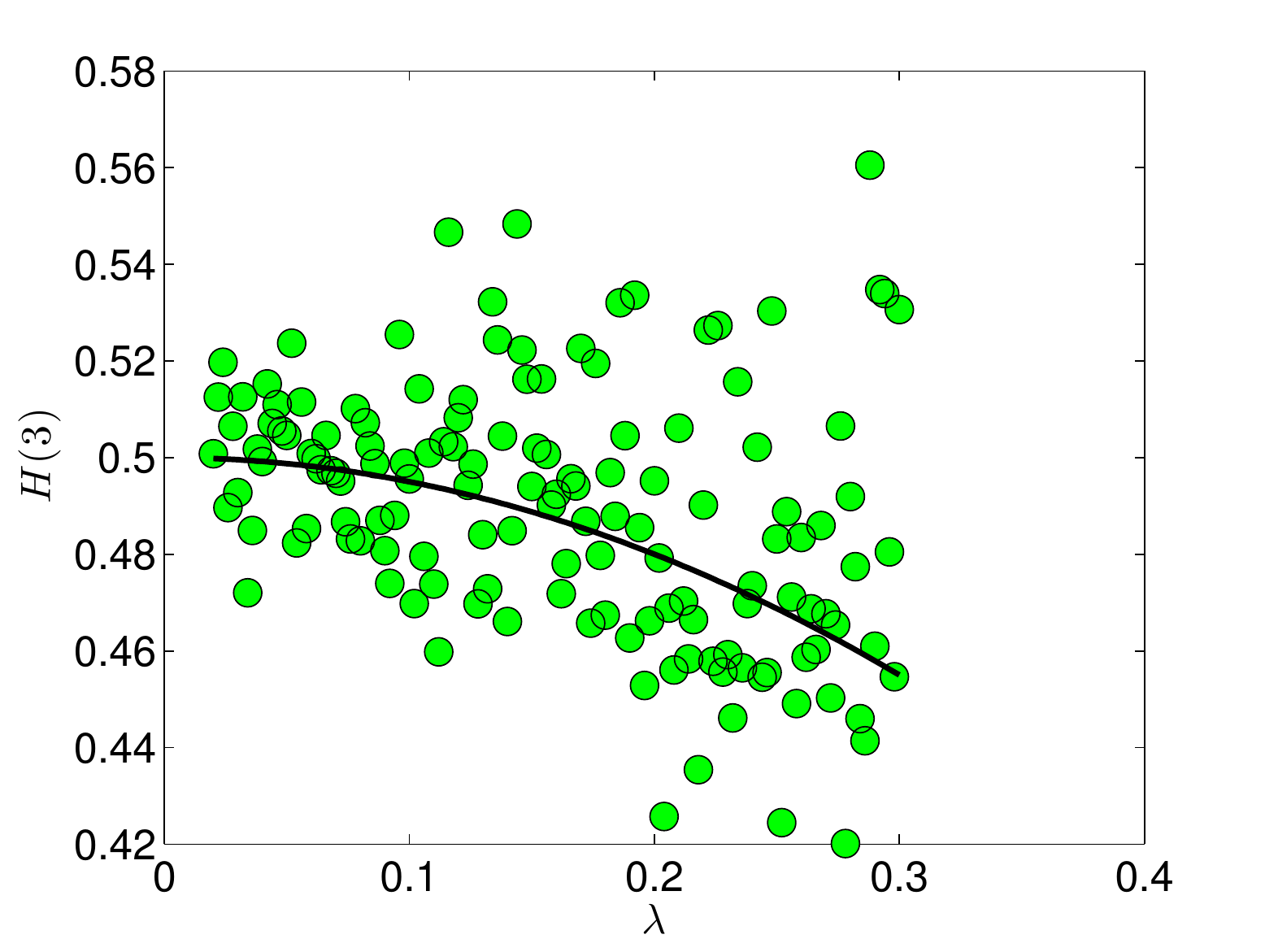}} }
\caption{\label{fig2} \textit{ We plot the $H(q)$ vs $\lambda$ for $q=1$ (left) and $q=3$ (right). The dots refer to the H(q)'s values obtained from the MRW-simulated time series with $\lambda$ ranging in [0.01,0.3]. The black solid lines are the theoretical relations (\ref{Hs}) for $q=1,3$. The GHE's have been obtained using \cite{matlabCode} with $\tau_{max}$ varying between $10$ and $30$, whereas $\Delta t$ has been set to $\Delta t=1$. } }
\end{figure}

The multifractal random walk (MRW) \cite{bacry2001multifractal} can be viewed as a stochastic volatility model constructed by taking the limit for $\Delta t\to 0$ of the process 
\begin{equation}\label{MRW}
X_{\Delta t}(t)=\sum_{k=1}^{t/\Delta t}\epsilon_{\Delta t}(k)\, e^{\omega_{\Delta t}(k)}
\end{equation}
with $\epsilon_{\Delta t}(k)$ a Gaussian white noise with variance $\sigma^2 \Delta t$ and $e^{\omega_{\Delta t}(k)}$ a stochastic volatility uncorrelated with $\epsilon$. By taking $\omega_{\Delta t}(k)$ as a stationary Gaussian process, we have log-normal volatility components. What distinguishes the limit $\Delta t\to 0$ of $X_{\Delta t}(t)$ from a Brownian motion is the choice of the auto covariance structure of the process $\omega_{\Delta t}(k)$, which is chosen, according to cascade-like processes \cite{arneodo1998analysis}, as 
\begin{equation}\label{covmrw}
\text{Cov}(\omega_{\Delta t}(k),\omega_{\Delta t}(k+h))=
\begin{dcases}
&\lambda^{2}\log\frac{T}{(1+h)\Delta t}, \quad h\leq T/\Delta t -1\\
&0, \quad \text{otherwise}.
\end{dcases}
\end{equation}
The logarithmic decay with lag $h$ of the auto covariance creates long memory in the process. This specification implicitly defines the integral scale $T$ and the intermittency coefficient $\lambda$. In order for the process $X_{\Delta t}$ to have finite variance in the limit $\Delta t\to 0$, one should impose $E (\omega_{\Delta t}(k))=-\text{Var}(\omega_{\Delta t}(k))=-\lambda^2 \ln(T/\Delta t)$ \cite{bacry2001multifractal}. $X(t)$ can be shown to obey self similarity exactly, i.e. for a time scale contraction $\tau'=s\tau$ ($s<1$) 
\begin{equation}\label{cascade}
X(t+\tau^{'})-X(t)=e^{\Gamma_{s}}[X(t+\tau)-X(t)],
\end{equation}
with $\Gamma_{s}$ a Gaussian random variable of variance $\text{Var}(\Gamma_{s})=-\lambda^2\ln(s)$ \cite{bacry2008continuous}. From equation (\ref{cascade}) one can show that \cite{muzy2010intermittency}
\begin{equation}\label{scalingMRW}
\mathbb{E}(|X(t+\tau)-X(t)|^q)=K_{q}\left(\frac{\tau}{T}\right)^{\zeta(q)},
\end{equation}
with $K_{q}$ a q-dependent factor and $\zeta(q)$ a non linear function of $q$, i.e. the process $X(t)$ is multi-scaling. It should be understood that the scaling (\ref{scalingMRW}) is exact only in the continuous time limit $\Delta t\to 0$. Nonetheless for $\Delta t < \tau$ one can recover good approximations of the scaling even when considering the discretized version $X_{\Delta t}(t)$.  
\newline The main appeal of this model for describing stock returns evolution lies in its ability to reproduce faithfully the most common stylised facts of financial markets: the hyperbolic decay of the volatility auto covariance function (\ref{covmrw}) as well as the heavy tails of the process increments. Indeed, as shown in \cite{bacry2008continuous}, the probability of observing increments larger than a certain value $x$ decays as a power law for large $x$:
\begin{equation}
\mathbb{P}\{|X(t+\tau)-X(t)|>x\}\sim x^{-\frac{1}{2\lambda^2}}.
\end{equation}
The parameter $\lambda$ controls therefore also the thickness of the tails of the returns distribution. The stationarity and the causal structure of this model make it also preferable to other multifractal models \cite{borland2005dynamics}. One of the most important features though, is that the multi-fractal spectrum of this model can be computed exactly to be \cite{bacry2001multifractal}
\begin{equation}\label{zetaspectrum}
\zeta_{q}=(q-q(q-2)\lambda^2)/2.
\end{equation}
The scaling function $\zeta (q)$ is therefore a parabola whose constant concavity only depends on $\lambda^2$. 
\newline Let us here start to look at the theoretical relation between the intermittency coefficient $\lambda$ and the GHE that can be established from the identity $\zeta_{q}=qH(q)$. Since the scaling exponents $H(q)$ are computed from the scaling of the empirical moments, one has for the log-normal MRW that $q$ is bounded from above by \cite{muzy2006extreme}
\begin{equation}\label{qlambdaln}
q\leq\frac{\sqrt{2}}{\lambda}. 
\end{equation}
Hence $\lambda$ and $q$ must be chosen jointly in such a way that the last inequality holds. As shown in Section \ref{sec3}, typical values of $\lambda$ obtained for financial data analysed in this study very rarely exceed the value of $0.3$, which is well within the bound of equation (\ref{qlambdaln}). By considering $q=1,2,3$ the scaling relation $\zeta_{q}=qH(q)$ for the MRW (using equation ($\ref{zetaspectrum}$)) reads 
\begin{align}
H(1)=&\frac{1}{2}(1+\lambda^{2}), \qquad  H(2)=\frac{1}{2}, \notag \\
H(3)=&\frac{1}{2}(1-\lambda^{2}). \label{Hs}
\end{align} 
\newline Values of the expressions (\ref{Hs}) for different $\lambda$'s are reported in Table \ref{firsttab} for $\lambda=0.2,0.25,0.3$. We have performed the following statistical study in order to compare theoretical predictions with GHE's computed on synthetic time series: for a given value of $\lambda$, we have simulated $1000$ MRW series and computed the corresponding GHE's for every synthetic series; then, assuming these GHE's are i.i.d., which is quite reasonable since every MRW simulation is independent from the others, we have obtained a sample distribution for the GHE's. From this distribution we have computed the $\{2.5\%,50\%,97.5\%\}$-quantiles which give the range of fluctuation of the GHE beyond which any observation can be deemed anomalous. We have performed this study for $q=\{1,2,3\}$, after having checked that the scaling doesn't hold for larger $q$'s. Results of this analysis for $\lambda=0.2,0.25,0.3$ and $T=1250,2500$ are reported in Table \ref{secondtab}. By comparing these values with those expected from the MRW model in Table \ref{firsttab} we can conclude that all values computed via the equations given in (\ref{Hs}) are compatible with the statistical confidence intervals computed on the synthetic time series. In Figure \ref{fig2} we plot the scaling exponents $H(1)$ and $H(3)$ computed on synthetic MRW time series with intermittency $\lambda\leq0.3$ together with the theoretical relations (\ref{Hs}) (black solid lines). The measured scaling exponents are found to be fluctuating around the theoretical relations. Although in this figure we show only the cases $q=1,3$, we have also verified that the case $q=2$ is consistent with the expected behaviour $H(2)=1/2$. 
\begin{landscape}
\begin{table}
\caption{\label{firsttab} Values of $H(q)$ for $q=\{1,2,3\}$ computed through the expressions in equation (\ref{Hs}) for $\lambda=0.2,0.25,0.3$.}
\centering
\begin{tabular}{c c c c}
\hline
 &$\lambda=0.2$ & $\lambda=0.25$ &$\lambda=0.3$ \\ \hline
 H(1) & 0.52 & 0.53 &  0.54\\ \hline
 H(2) & 0.50 & 0.50 & 0.50 \\  \hline
 H(3) & 0.48 & 0.47 & 0.45 \\  \hline
\end{tabular}
\end{table}
\begin{table}
\caption{\label{secondtab} We report numerical results of the $\{2.5\%,50\%,97.5\%\}$-quantiles obtained from the distributions of the GHE's computed on several synthetic MRW time series for $q=1,2,3$, for $\lambda=0.2,0.25,0.3$ and $T=1250,2500$. We can appreciate dependence of the quantiles both on $\lambda$ and $q$.  }
\centering
\begin{tabular}{llccc}
\hline\hline
& & \multicolumn{3}{c}{$\{2.5\%,50\%,97.5\%\}-\text{quantiles}$}\\
\cline{3-5}
& & H(1) & H(2) & H(3) \\ \hline \hline
\multirow{2}{*}{$\lambda=0.2$} & T=1250 & $\{0.4782,0.5110,0.5427\}$ & $\{0.4592,0.4986,0.5381\}$ & $\{0.4326,0.4838,0.5421\}$ \\
& T=2500 & $\{0.4768,0.5101,0.5450\}$& $\{0.4562,0.4965,0.5407\}$ & $\{0.4284,0.4807,0.5454\}$ \\ \hline\hline
\multirow{2}{*}{$\lambda=0.25$} & T=1250& $\{0.4844,0.5167,0.5502\}$& $\{0.4527,0.4976,0.5415\}$ & $\{0.4029,0.4739,0.5488\}$ \\
& T=2500 & $\{0.4843,0.5180,0.5517\}$& $\{0.4530,0.4988,0.5428\}$& $\{0.4038,0.4774,0.5479\}$\\ \hline\hline
\multirow{2}{*}{$\lambda=0.3$} & T=1250& $\{0.4900,0.5249,0.5616\}$& $\{0.4374,0.4962,0.5571\}$& $\{0.3642,0.0458,0.5597\}$ \\
& T=2500 & $\{0.4878,0.5253,0.5626\}$ & $\{0.4398,0.4968,0.5558\}$ & $\{0.3743,0.4653,0.5609\}$  \\ 
\hline\hline
\end{tabular}
\end{table}
\end{landscape}
It has been shown that modifications of the log-normal model, where residuals $\epsilon_{\Delta t}(t)$ are Gaussian, can account better for the fat tails observed in empirical data \cite{muzy2006extreme}. One may first consider residuals $\epsilon_{\Delta t}(t)$ to be Student t distributed, that is (we drop the subscript $\Delta t$ for readability)
\begin{equation}
P(\epsilon)=\frac{1}{\sqrt{\pi}}\frac{\Gamma\left(\frac{1+\nu}{2}\right)}{\Gamma(\frac{\nu}{2})}\frac{a^{\nu}}{(\epsilon^2+a^2)^{\frac{1+\nu}{2}}},
\end{equation}
where $\nu$ is the number of degrees of freedom, $\Gamma$ is the Euler gamma function and the parameter $a$ is related to the variance $\sigma^2$ via $\sigma^2=a^2/(\nu-2)$ \cite{bouchaud2003theory}. The Student t distribution is known to better fit stock returns tails, if one considers $\nu=4$ or $5$ \cite{bouchaud2003theory}. In order to let the model have fatter tails one can also act on the unconditional distribution of the $\omega_{\Delta t}$; in \cite{muzy2006extreme} indeed, it was pointed out that a cascade model where $\omega_{\Delta t}$ follows a gamma law can account for empirical observations better than a normal law. The pdf of $\omega_{\Delta t}$ is given in this case by (again the subscript $\Delta t$ is dropped) 
\begin{equation}\label{loggamma}
P(\omega)=\frac{1}{\Gamma(k)\theta^{k}}\omega^{k-1}e^{-\omega/\theta},
\end{equation}
where $k$ and $\theta$ are respectively shape and scale parameters. We will use these two modifications of the log-normal MRW in Sections \ref{secT} and \ref{secGamm}.   
\section{\label{sec3} Analysis of empirical data: time-varying multifractality}
\subsection{Statistical testing procedure}\label{StatTest}
The hypothesis we want to test in this paper is whether stock returns exhibit multifractal properties consistent with a multifractal model whose intermittency coefficient $\lambda$ is constant. First of all one needs to check that the data analysed show indeed multifractal properties, a feature which is encoded in the concavity of the estimated scaling function $\zeta_{q}$. Once the scaling has been checked we have performed the following statistical testing procedure: after having estimated the MRW parameters from the empirical data we simulated 1000 MRW synthetic series of length $n=1250$ with the parameters obtained from the empirical time series (details in Section \ref{secEmp}). On each of these synthetic series we have computed the multifractality proxy
\begin{equation}
\Delta H^{w}(1,2)=H^{w}(1)-H^{w}(2)
 \end{equation}
via the weighted generalised Hurst exponent method \cite{matlabCode} with damping coefficient equal to $415$ days, chosen in agreement with previous analysis carried out in \cite{morales2012dynamical,pozzi2012exponential}. The scaling exponents $H^{w}(1)$ and $H^{w}(2)$ have been computed as average of several fits of equation (\ref{scaling2}) with the scale $\tau\in[1,\tau_{max}]$, with $\tau_{max}$ varied between $10$ and $30$. Computing the scaling exponents for each of the simulated series we have obtained a distribution of the measured $\Delta H^{w}(1,2)$'s. We have then computed dynamically on $1250$ days rolling windows, $\Delta H^{w}_{t}(1,2)$ on the empirical time series and compared its fluctuations in time with the extreme quantiles of the distributions obtained from synthetic series. We have then looked at the percentage of fluctuations falling outside of the confidence intervals provided by the extreme quantiles. A high quantile-exceedance rate of the empirical $\Delta H^{w}_{t}(1,2)$ would suggest that the fluctuation in multifractality envisaged by the model are smaller than those observed on empirical data. If this is the case, these findings would suggest that a time-varying intermittency coefficient may be responsible for this feature observed in financial time series. 
\subsection{Empirical analysis}\label{secEmp}
\begin{figure}
\centering
\includegraphics[width=0.7\columnwidth]{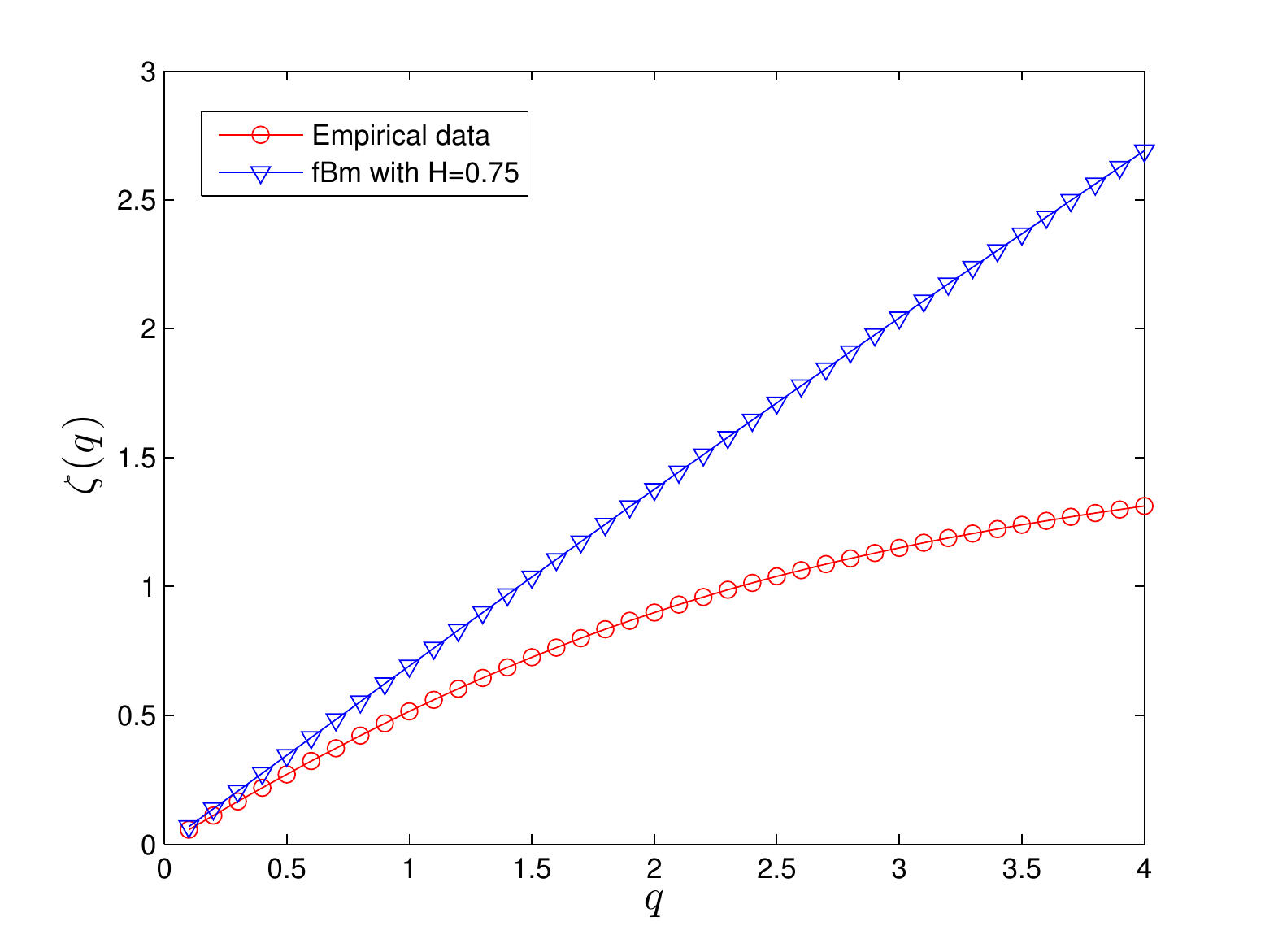}
\caption{\label{fig1} \textit{The scaling function for Microsoft Corp daily data is compared to that of a notorious unifractal model, i.e. the fractional Brownian motion. The choice of $H=0.75$, arbitrary, means we are considering a process with long memory although unifractal. } }
\end{figure}

\begin{figure}[h!]
\begin{tabular}{cc}
\includegraphics[width= 0.5\columnwidth]{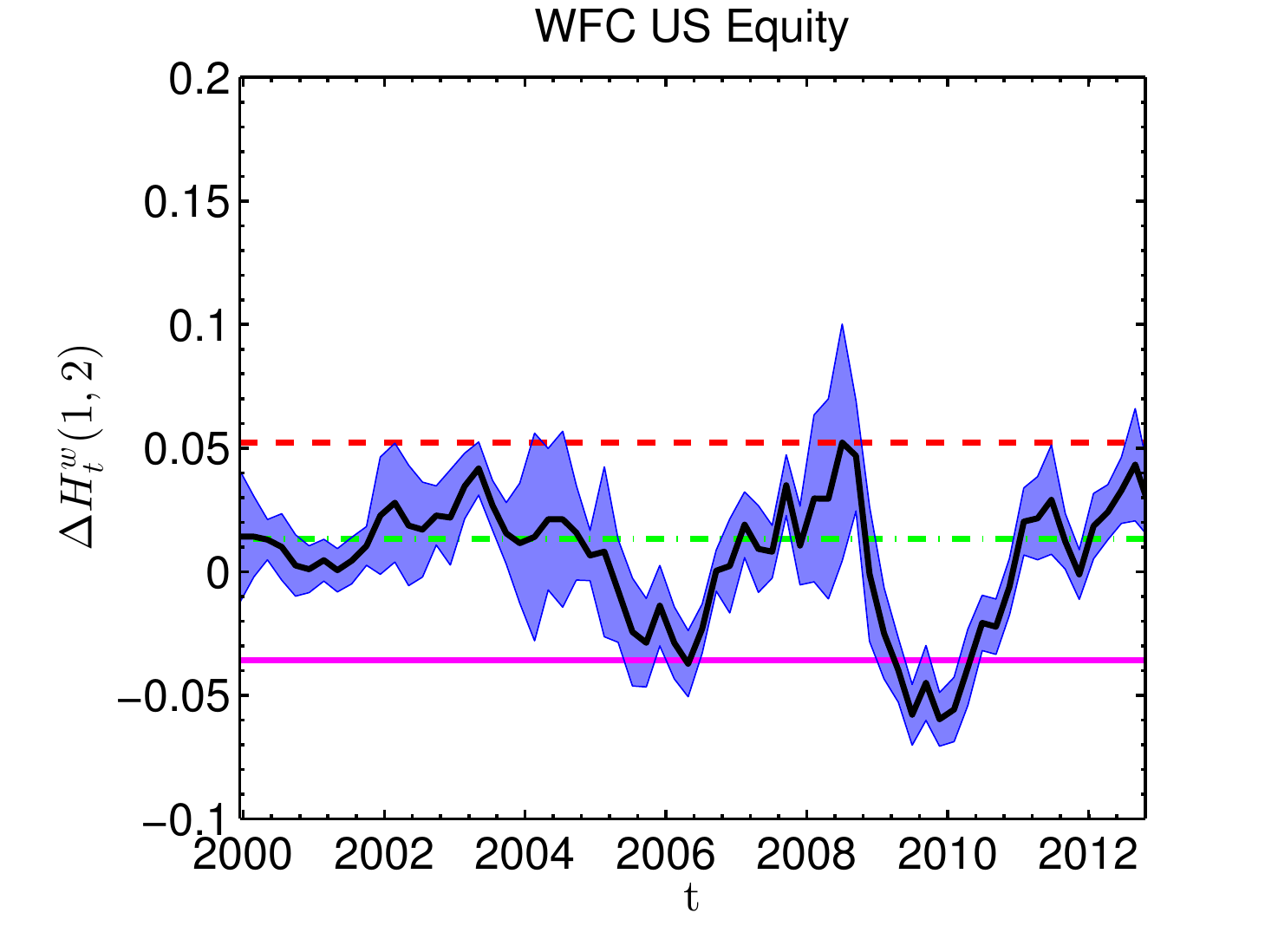} &
\includegraphics[width= 0.5\columnwidth]{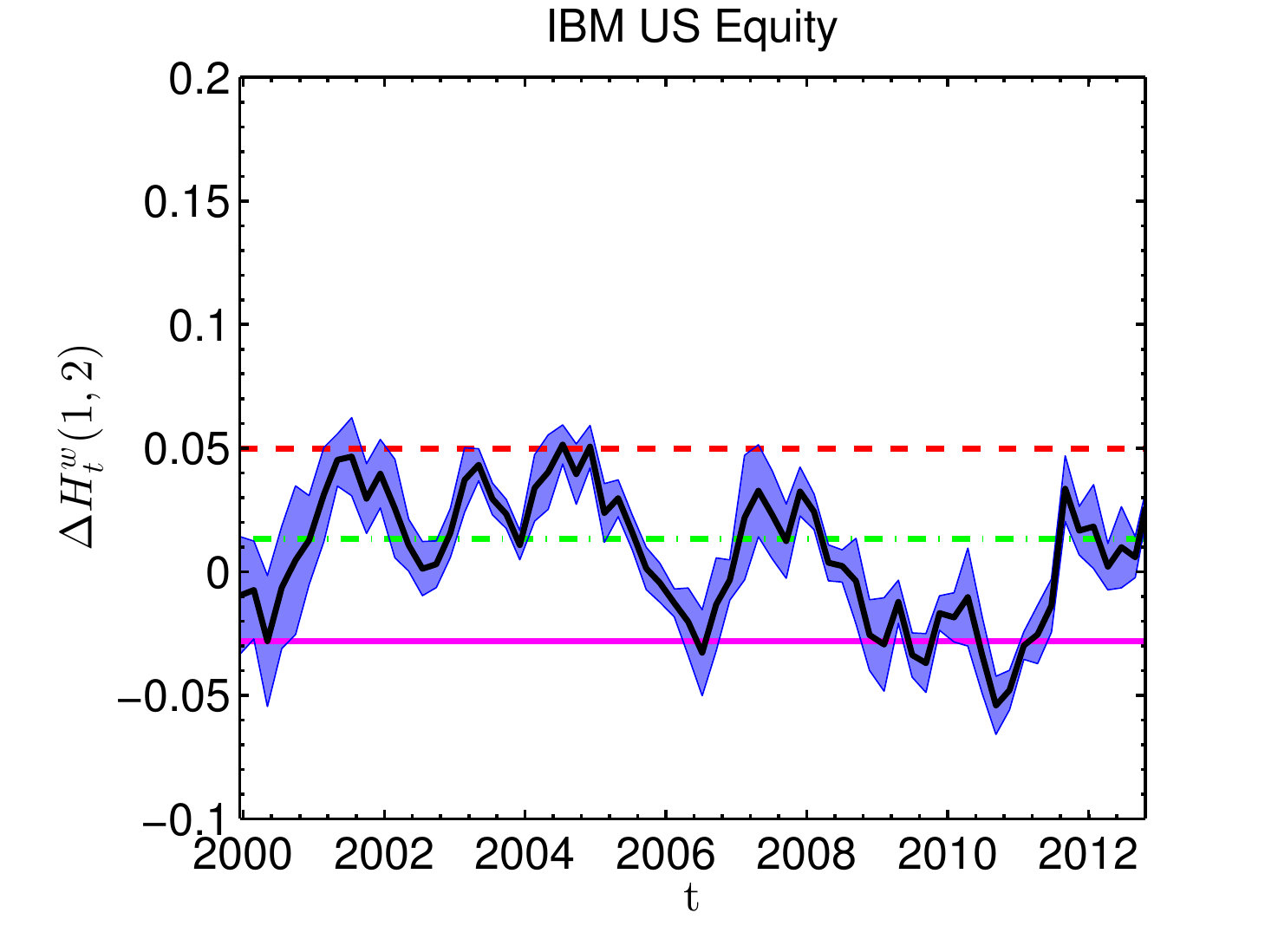} \\
\includegraphics[width= 0.5\columnwidth]{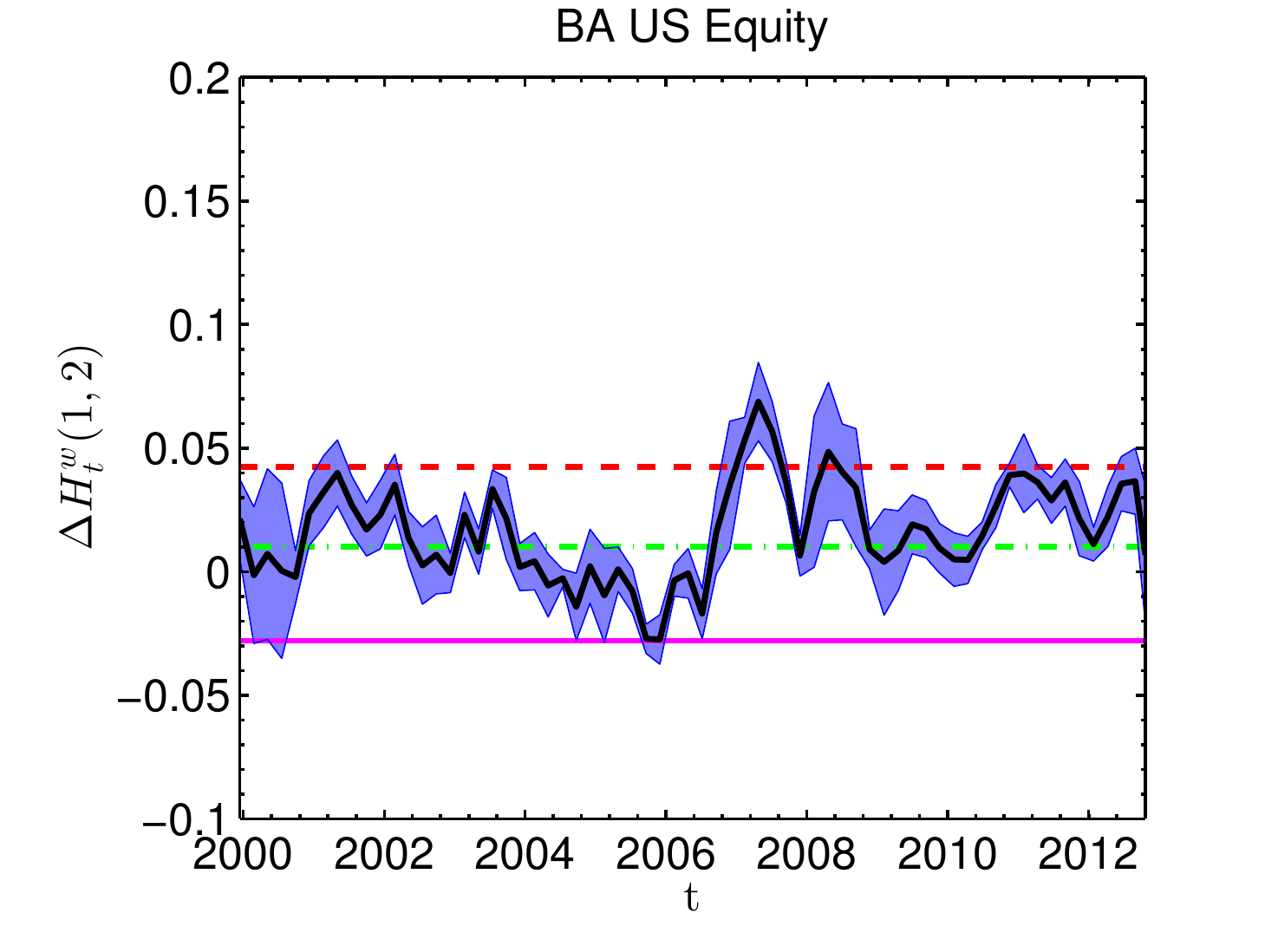} &
\includegraphics[width= 0.5\columnwidth]{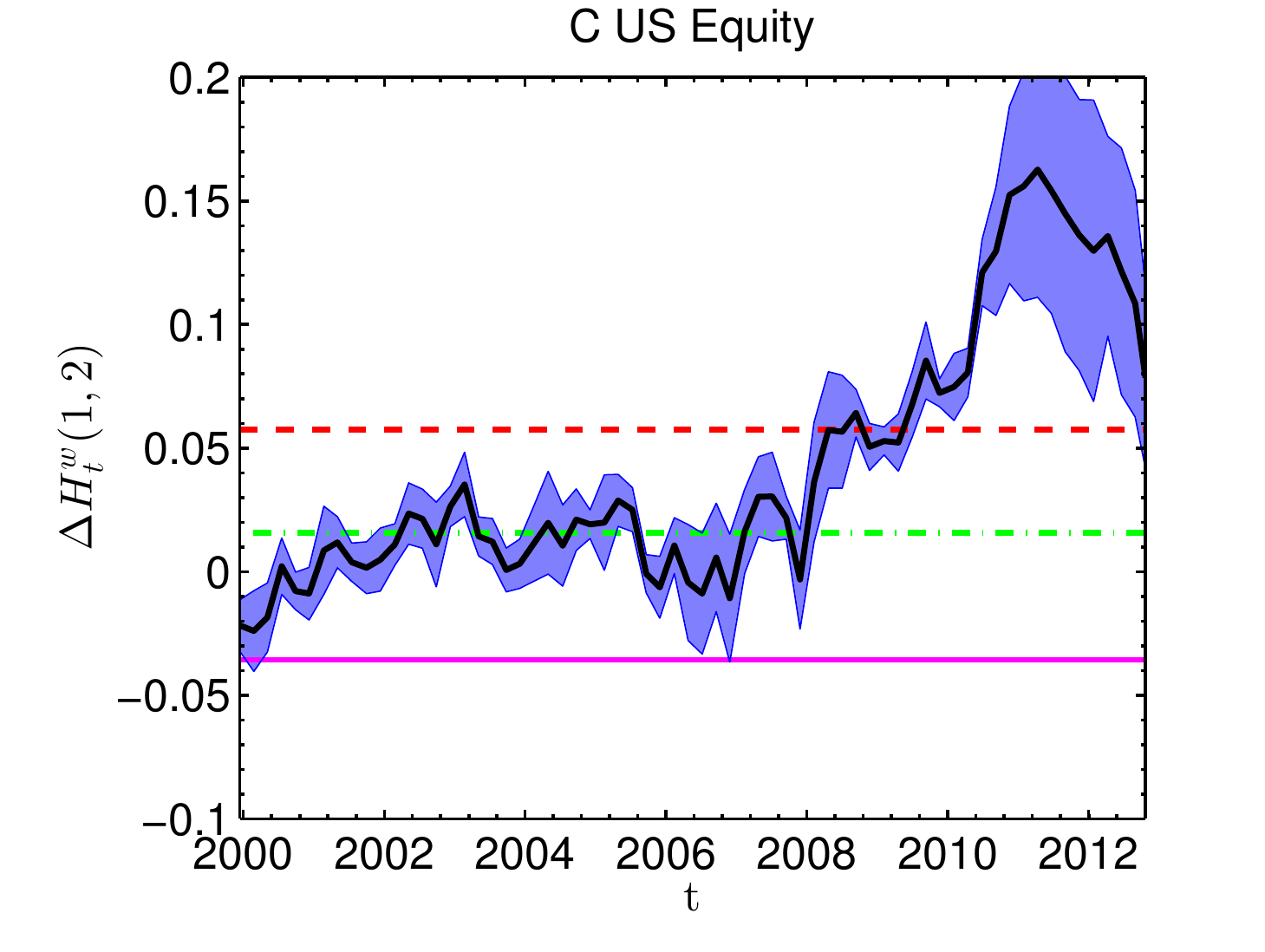} \\
\includegraphics[width= 0.5\columnwidth]{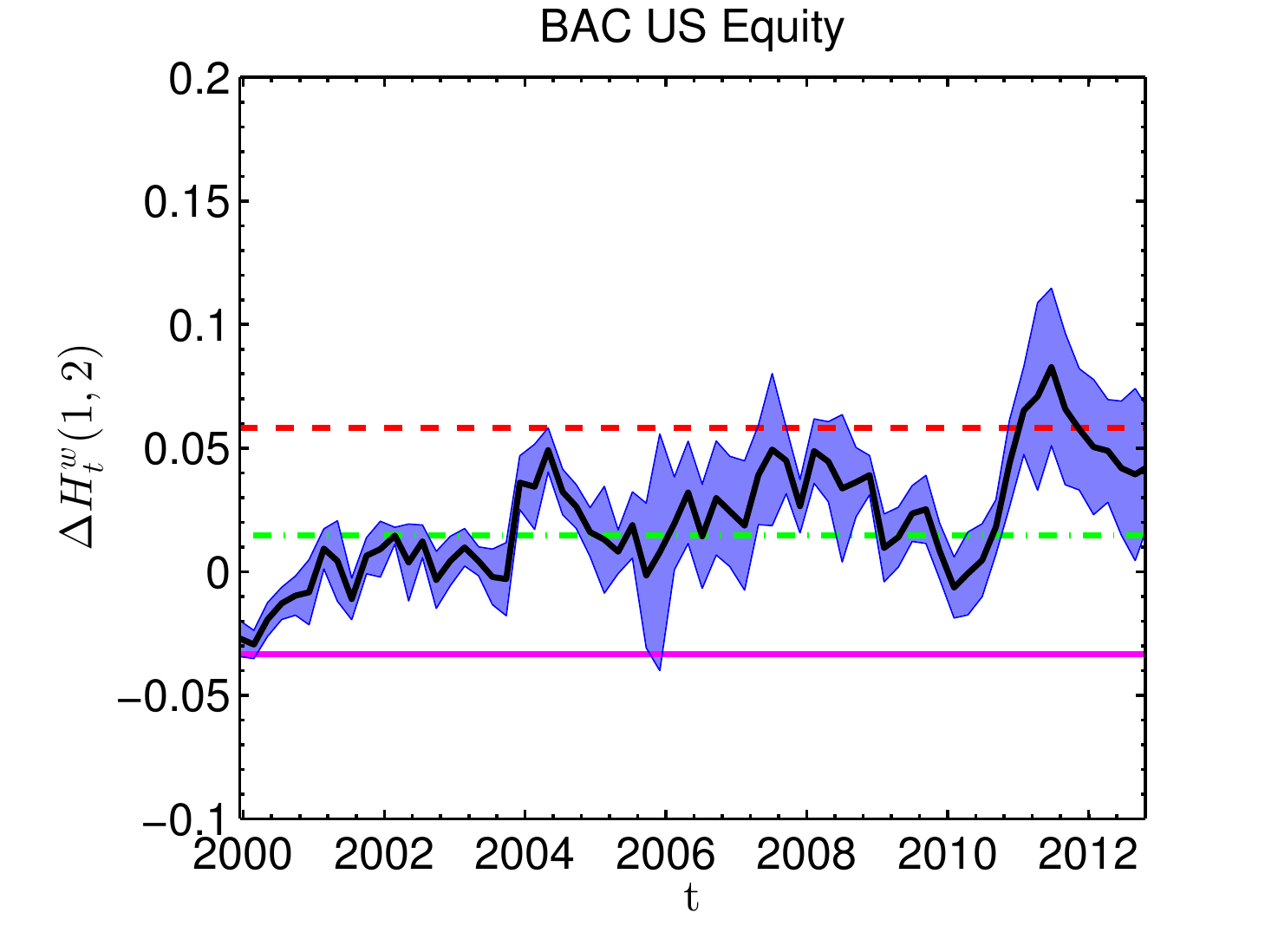} &
\includegraphics[width= 0.5\columnwidth]{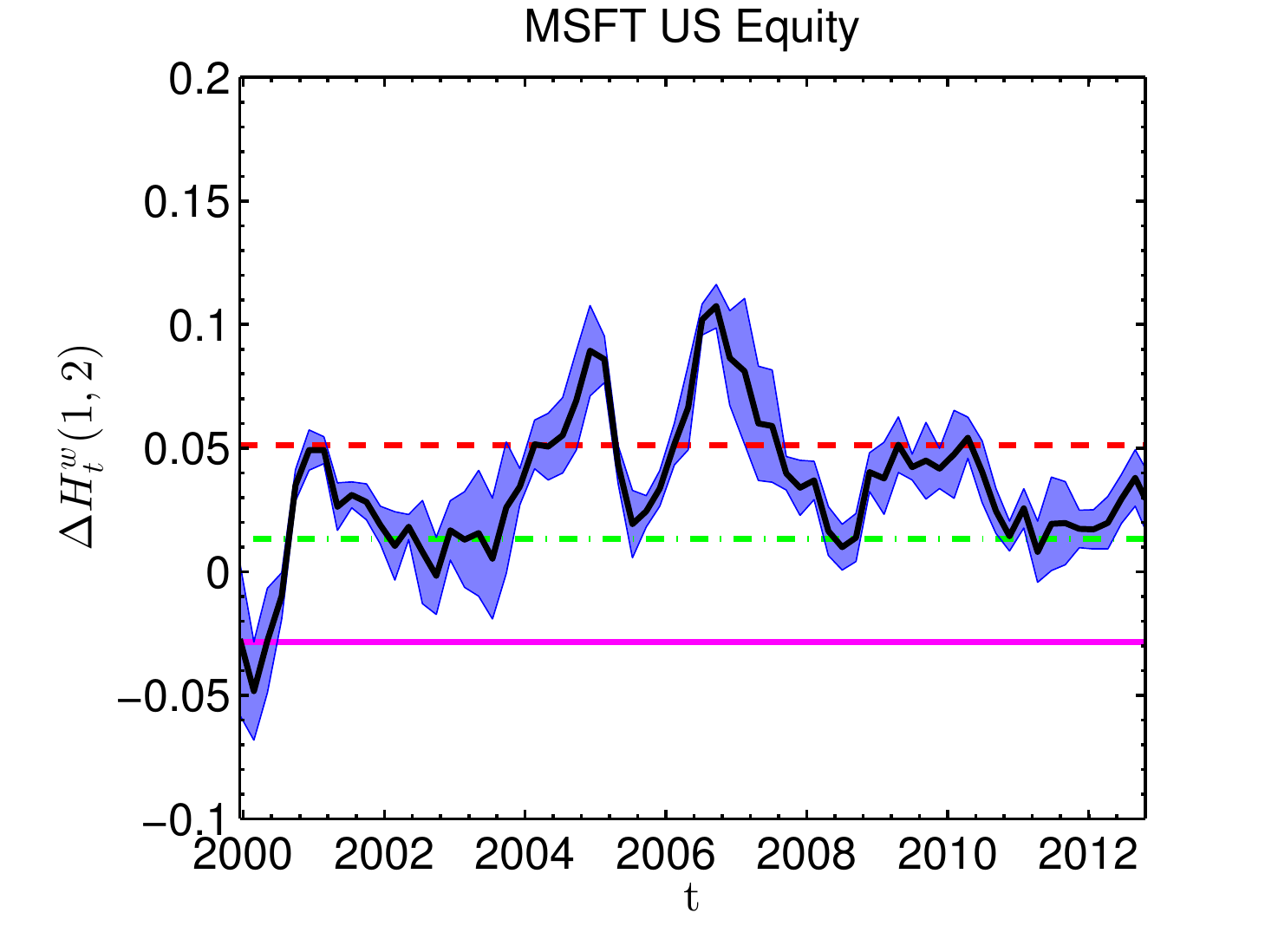}
\end{tabular}
\caption{\label{fig6}
\textit{We plot the dynamical evolution of $\Delta H^{w}(1,2)$ ($\Delta H^{w}_{t}(1,2)$) (black thick line) for six stocks in the time period 01-01-1995 to 22-10-2012. $\Delta H^{w}_{t}(1,2)$ is evaluated on moving overlapping time windows of 1250 days, with a shift of 100 days. The first time window includes data from 1995 to 2000 and the corresponding $\Delta H^{w}_{t}(1,2)$ is plotted at the end of the period. The horizontal lines represent the 2.5\% (red dashed line), 50\% (green dot-dashed line) and 97.5\% (magenta continuous line) quantiles extracted from the distribution of $\Delta H^{w}(1,2)$ estimated from 1000 realisations of MRW with parameters extracted from the data. The blue shaded area represents the error on $\Delta H^{w}_{t}(1,2)$, computed, for each t, as the standard deviation on $\Delta H^{w}(1,2)$. } }
\end{figure}

The statistical test described in Section \ref{StatTest} has been performed on a set of $340$ daily stock prices quoted in the NYSE in the period ranging from 01-01-1995 to 22-10-2012. The data were provided by Bloomberg. We have first of all verified that the scaling functions of the data are indeed not linear in $q$. As an example, we plot in Figure \ref{fig1} the scaling function of the daily stock returns of Microsoft Corp compared to that of a fractional Brownian motion with Hurst parameter $H=075$.\footnote{The choice of $H$ is arbitrary, as the linear scaling is expected for all $H$.}  This scaling has been checked to hold for the whole set of stocks analysed. 
\newline The parameters of the MRW can be extracted from the empirical data through the behaviour of the log-volatility auto covariance function. By identifying empirical returns with the increments of the MRW, using equation (\ref{MRW}) one can write 
\begin{equation}\label{retomega}
\log (|r_{t,\tau}|)=\omega_{\Delta t}(k)+\log |\epsilon_{\Delta t}(k)|.
\end{equation}
It has been shown \cite{muzy2000modelling} that for financial time series, the function 
\begin{equation}
C(h)=\text{Cov}(\log |r_{t+h,\tau}|, \log|r_{t,\tau}|)
\end{equation}
exhibits a slow decay with the lag $h$. By putting together the last two equations we see that $C(h)$ is proportional to the log-volatility auto covariance function plus a term uncorrelated in the lag $h$, i.e. 
\begin{equation}\label{ChLambda}
C(h)\sim\lambda^{2}\log \frac{T}{(1+h)\Delta t}.
\end{equation}
The intermittency coefficient $\lambda$ and the integral scale $T$ can therefore be estimated from linear fits of equation (\ref{ChLambda}) in log-linear scale. As already established in previous studies \cite{muzy2010intermittency}, the estimation of $\lambda$ through this method yields much more reliable results than an estimation based on the variogram of the $\omega_{\Delta t}(k)$ as a function of the scale $\Delta t$ (see \cite{muzy2010intermittency} for details and comparison of the two methods). In the whole procedure we fix the return scale at $\Delta t=1$ day. These simulations of MRW synthetic series have been performed using a Fast Fourier Transform of the auto covariance function (\ref{covmrw}).  
\begin{landscape}
\begin{table}
\caption{\label{fourthtab} \textit{In this table we report, for 13 different companies, the parameters $\lambda,T,\sigma$ of the MRW obtained from empirical data. The fourth column shows the quantiles of the distributions of $\Delta H^{w}(1,2)$ computed from 1000 realisations of MRW synthetic series of length $1250$, simulated with the corresponding set of parameters. The fifth column shows the percentage of dynamical $\Delta H^{w}(1,2)$'s, computed on 1250 time steps rolling windows on the empirical time series, falling outside the confidence interval given in the fourth column. }}
\begin{tabular}{lccccc}
& $\lambda$ & T (days) & $\sigma$ & Quantiles & Exceedances percentage \\ \hline \hline
Boeing Corp & 0.1228 & 1260  & 1.0651 & $\{-0.0205,0.0039,0.0265\}$ & 8 \%  \\ \hline\hline
Microsoft Corp &  0.1342 & 739 & 0.021 &$\{-0.03341,0.01067,0.04675\}$ & 33 \%\\ \hline\hline
PNC Financials & 0.1418 & 28201 & 0.025 & $\{-0.03191,0.0098,0.0349 \}$ & 7 \%\\ \hline\hline
Sara Lee Corp & 0.1335 & 3746 & 0.019 & $\{-0.03381,0.00948,0.04524\}$ &  31.5\%\\ \hline\hline
Rowan Cos INC & 0.2046 & 1080 & 0.7992 & $\{-0.0215,0.0066,0.0296\}$ & 11.4\%\\ \hline\hline
IBM Corp & 0.2363 & 816 & 1.7811 & $\{-0.0214,0.0068, 0.0310\}$ &  14\%\\ \hline\hline
Wells Fargo & 0.3010 & 631& 0.5923 & $\{-0.0237, 0.0134, 0.0433\}$ &  15\%\\ \hline\hline
American Express & 0.1340 &  1520 & 0.8356 & $\{-0.0212,0.0125,0.0429\}$ &  29\%\\ \hline\hline
General Motors Corp & 0.1727 & 440 & 1.0291 &$\{-0.0211, 0.0063, 0.0278 \}$ &  41\%\\ \hline\hline
Citigroup & 0.2196 & 568 & 6.4585 & $\{ -0.0202, 0.0146, 0.0458\}$ &  39\%\\ \hline\hline
JPMorgan Chase & 0.2950 & 534 & 0.9792 & $\{-0.0219,0.0097,0.0356\}$ & 27 \% \\ \hline\hline
Dominion Resources INC & 0.1738 & 754 & 0.013 & $\{-0.0317,0.01372,0.05411\}$ & 10 \% \\ \hline\hline
Morgan Stanley & 0.2528 & 538 & 1.1185 & $\{-0.0221,0.0079,0.0333\}$ &14 \% \\ \hline\hline
\end{tabular}
\end{table}
\end{landscape}
We have estimated the MRW parameters for each empirical time series. Then, for each set of parameter we have simulated 1000 MRW synthetic series of 1250 time steps and on each series computed $\Delta H^{w}(1,2)$. The distribution of these $\Delta H^{w}(1,2)$ for each set of parameters is used to estimate the $\{2.5\%,50\%,97.5\%\}$-quantiles. Values of the parameters obtained for different stocks together with the  corresponding quantiles obtained from the distributions of the $\Delta H^{w}(1,2)$'s obtained from the simulated MRW series are reported in Table \ref{fourthtab}. In Table \ref{fourthtab} we also report the percentage of empirical $\Delta H^{w}_{t}(1,2)$ computed on rolling windows of length $1250$ exceeding the quantiles, defined as 
\begin{equation}
\text{Exceedance}=\frac{1}{N}\left(\sum_{t=1}^{N}\mathbbm{1}_{\{\Delta H_{t}^{w}(1,2)>Q^{w}_{97.5\%}\}}+\sum_{t=1}^{N}\mathbbm{1}_{\{\Delta H_{t}^{w}(1,2)<Q^{w}_{2.5\%}\}}\right),
\end{equation}
where $N$ is the number of time windows, $Q^{w}_{97.5\%}$ and $Q^{w}_{2.5\%}$ are respectively the $97.5\%$ and $2.5\%$ quantiles of the $\Delta H^{w}(1,2)$ distribution and $\mathbbm{1}_{\{x\}}$ is the indicator function which is $1$ if the condition $x$ is enforced and $0$ otherwise.
The rather high number of points falling outside the confidence intervals confirm systematically that empirical data do not agree with the hypothesis of constant multifractality. As one can appreciate from the quantiles reported in Table \ref{fourthtab}, there is no correlation between the number of quantile-exceeding $\Delta H^{w}_{t}(1,2)$ and $\lambda$. This fact tells us that a simple underestimation of $\lambda$ is unlikely to be the cause of the high quantile exceedance rate.    
In Figure \ref{fig6} we show the results of this study for several daily stock prices, where $\Delta H^{w}_{t}(1,2)$ is shown to exceed the quantiles many times.   
\newline We must remark that the values obtained for the correlation lengths T are very large compared to the time series lengths, a feature which is nonetheless commonly observed in multi-timescale volatility models \cite{borland2005multi}. 
\begin{figure}
\centering
\includegraphics[width=0.7\columnwidth]{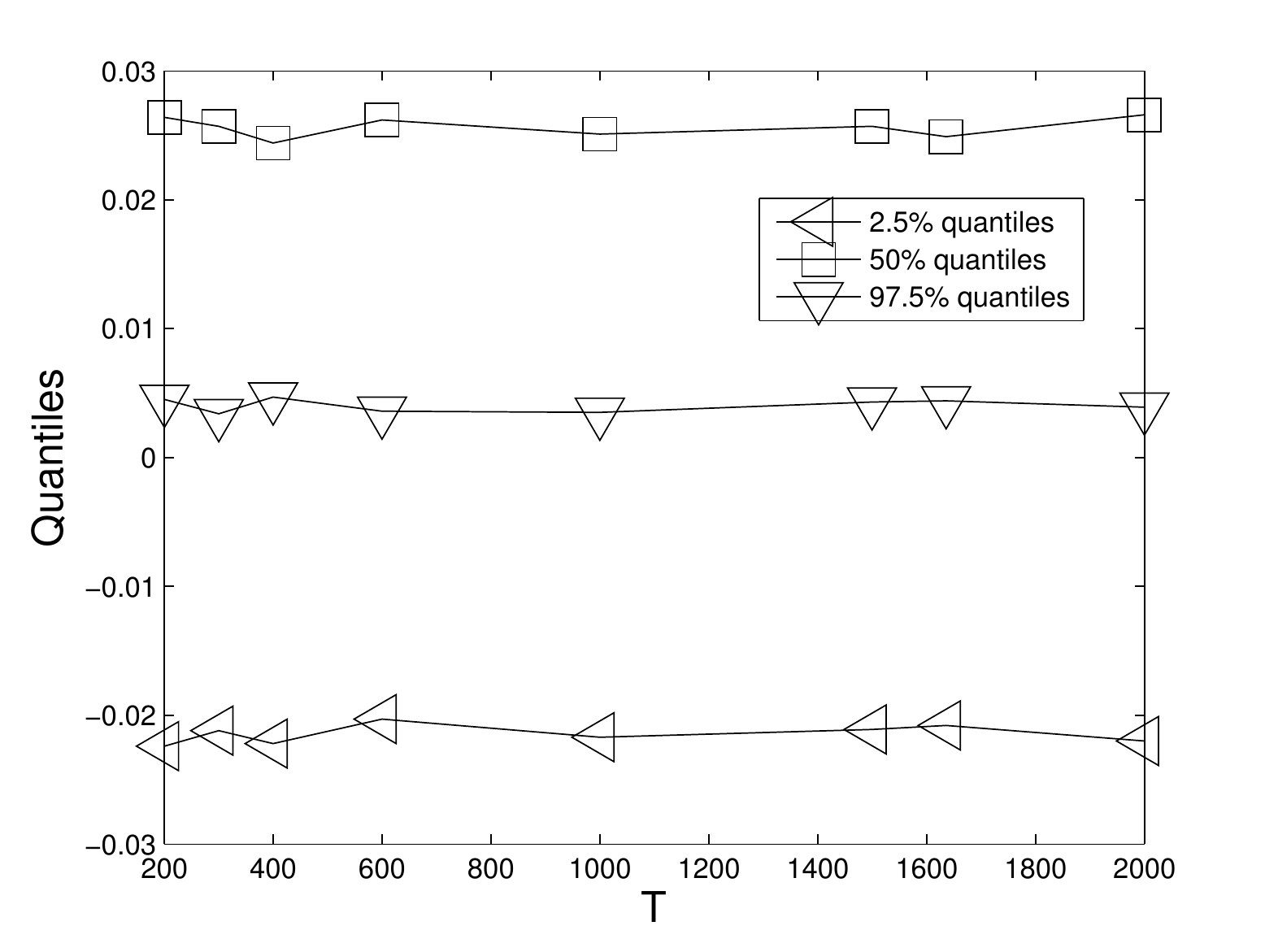}
\caption{\label{figT} \textit{We plot the quantiles of the $\Delta H^{w}(1,2)$ distribution as function of the integral scale $T$. We have fixed $\lambda=0.2$ and computed the $\Delta H^{w}(1,2)$ quantiles on synthetic series simulated with integral scales $T=200,300,400,600,1000,1500,1650,2000$. }}
\end{figure}
However, multifractality is expected to be proportional only to $\lambda^{2}$ and thus the values of $\Delta H^{w}(1,2)$ that we have measured should not be affected by the true $T$ of the series. This has been checked by computing the quantiles of the distributions of $\Delta H^{w}(1,2)$ simulated with a fixed $\lambda$ and varying the integral scale $T$. As one can appreciate in Figure \ref{figT} the quantiles do not vary when the integral scale of the simulated series is changed but $\lambda$ is kept constant. 
\begin{figure}
\includegraphics[width=\columnwidth,height=8 cm]{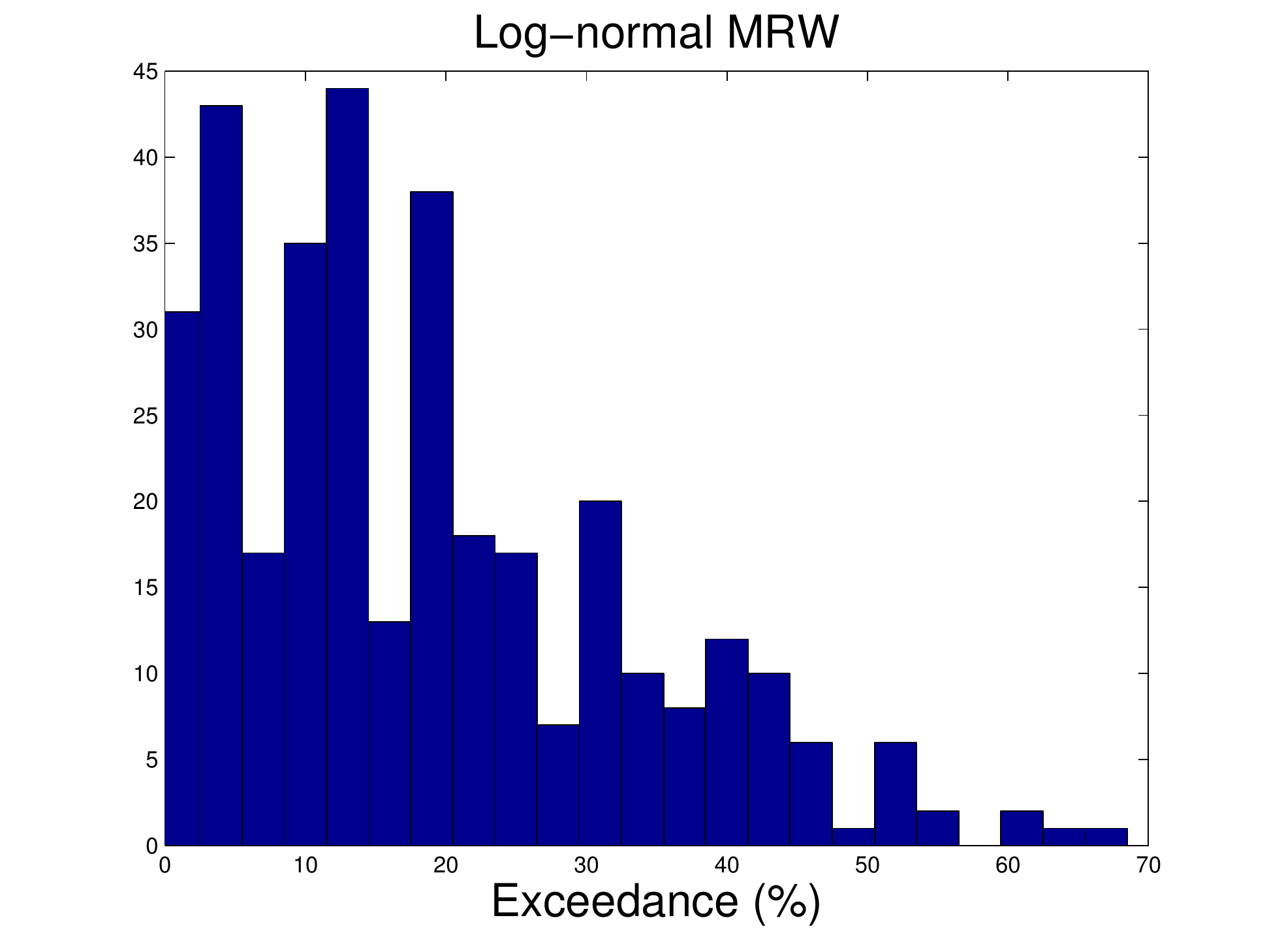}
\caption{\label{figHistograms}
\textit{Histogram of the observed exceedance percentages observed on the whole set of stocks when the benchmark model is the log-normal MRW. On the y-axes is the number of stocks exhibiting the percentage of quantile-crossings (given in the x-axis).
} }
\end{figure}
\newline We have performed this statistical test for the set of 340 stocks listed in the NYSE. In particular, for each empirical time series, we have compared the fluctuations of empirical $\Delta H_{t}^{w}(1,2)$ with the confidence intervals obtained from the MRW series simulated with the parameters estimated from the corresponding stock. The histogram in Figure \ref{figHistograms} shows the distribution of the observed percentage of exceedances on this data set. It is remarkable that a conspicuous number of stocks shows a relevant number of $\Delta H^{w}_{t}(1,2)$'s falling outside the $95\%$ confidence interval: the majority of stocks exhibits quantile crossings above $10\%$ and we find even some percentages above $60\%$. This histogram shows that globally the hypothesis of a log-normal MRW model with constant intermittency looks like a poor approximation of empirical time series behaviour. 
\newline The conclusion one draws from these observations is that our proxy of multifractality appears to defy the hypothesis of a constant multifractal behaviour, as in the setup of the MRW model, as the majority of stocks shows time variations of $\Delta H^{w}(1,2)$ which are not in agreement with the null hypothesis of constant multifractal behaviour. As discussed above, a time varying multifractality requires, through equation (\ref{zeta_Hurst}) and (\ref{zetaspectrum}), a time varying $\lambda$. The observations reported here suggest that the scaling properties of financial time series may vary over time because of the reflection of complex and varying economic constraints. However, one needs to pay attention to the fact that the log-normal MRW with Gaussian residual may be a poor approximation for the empirical observations of the return distribution tails. In the next two sections we show how the same analysis performs if we consider a model with thicker tails. 
\subsection{The effect of Student t residuals}\label{secT}
\begin{figure}
\begin{tabular}{cc}
\includegraphics[width= 0.5\columnwidth]{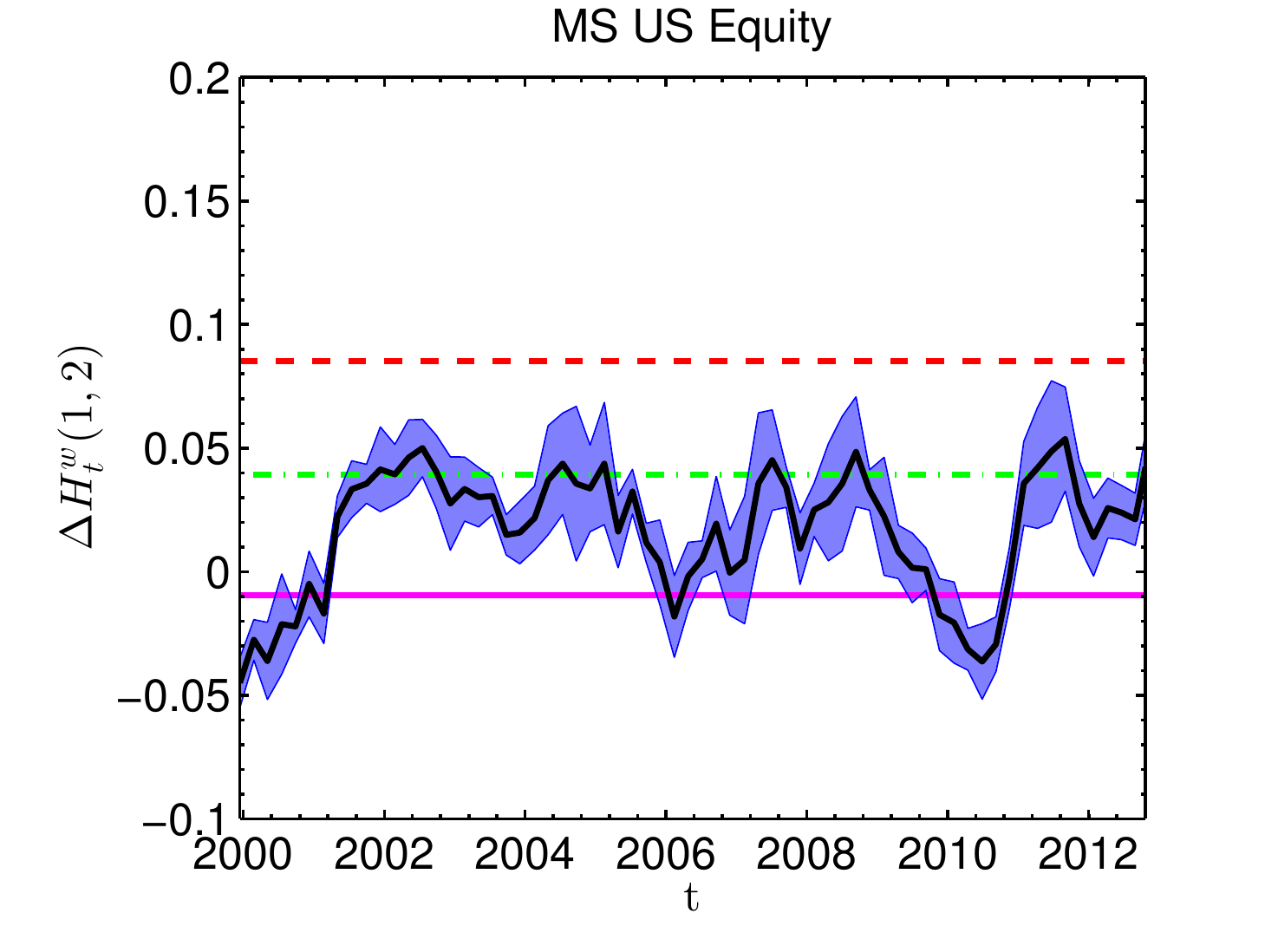} &
\includegraphics[width= 0.5\columnwidth]{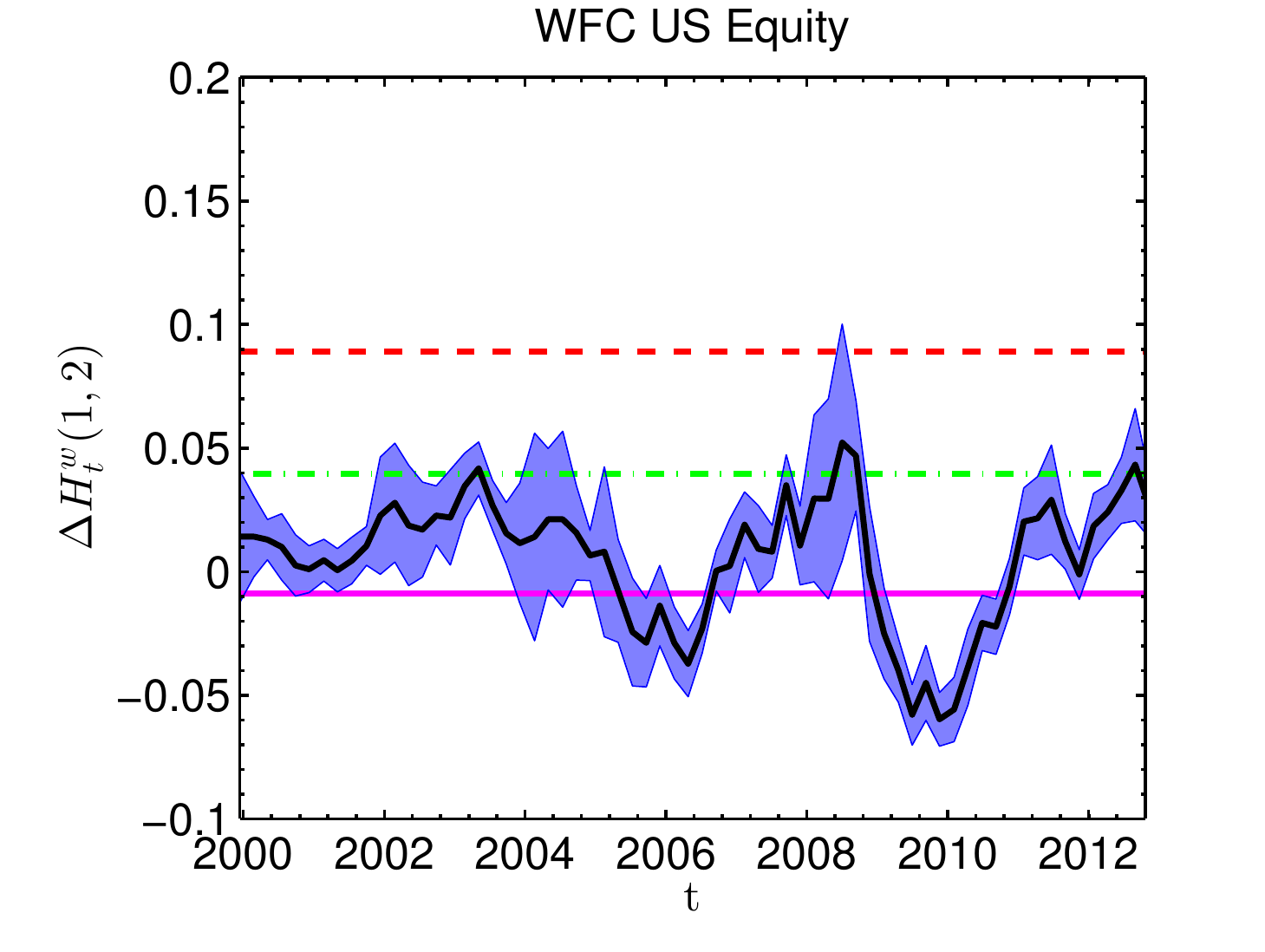} \\
\includegraphics[width= 0.5\columnwidth]{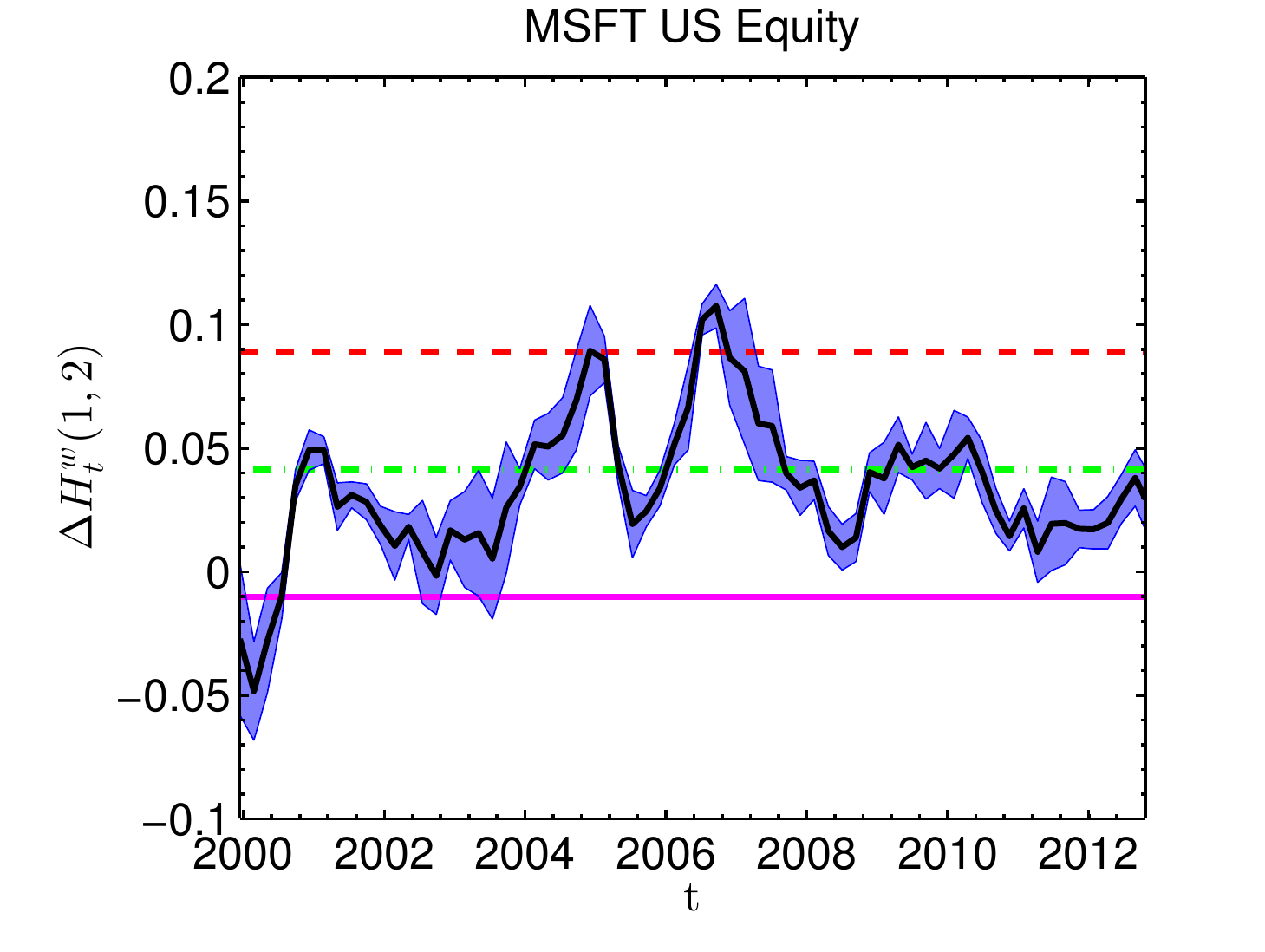} &
\includegraphics[width=0.5 \columnwidth]{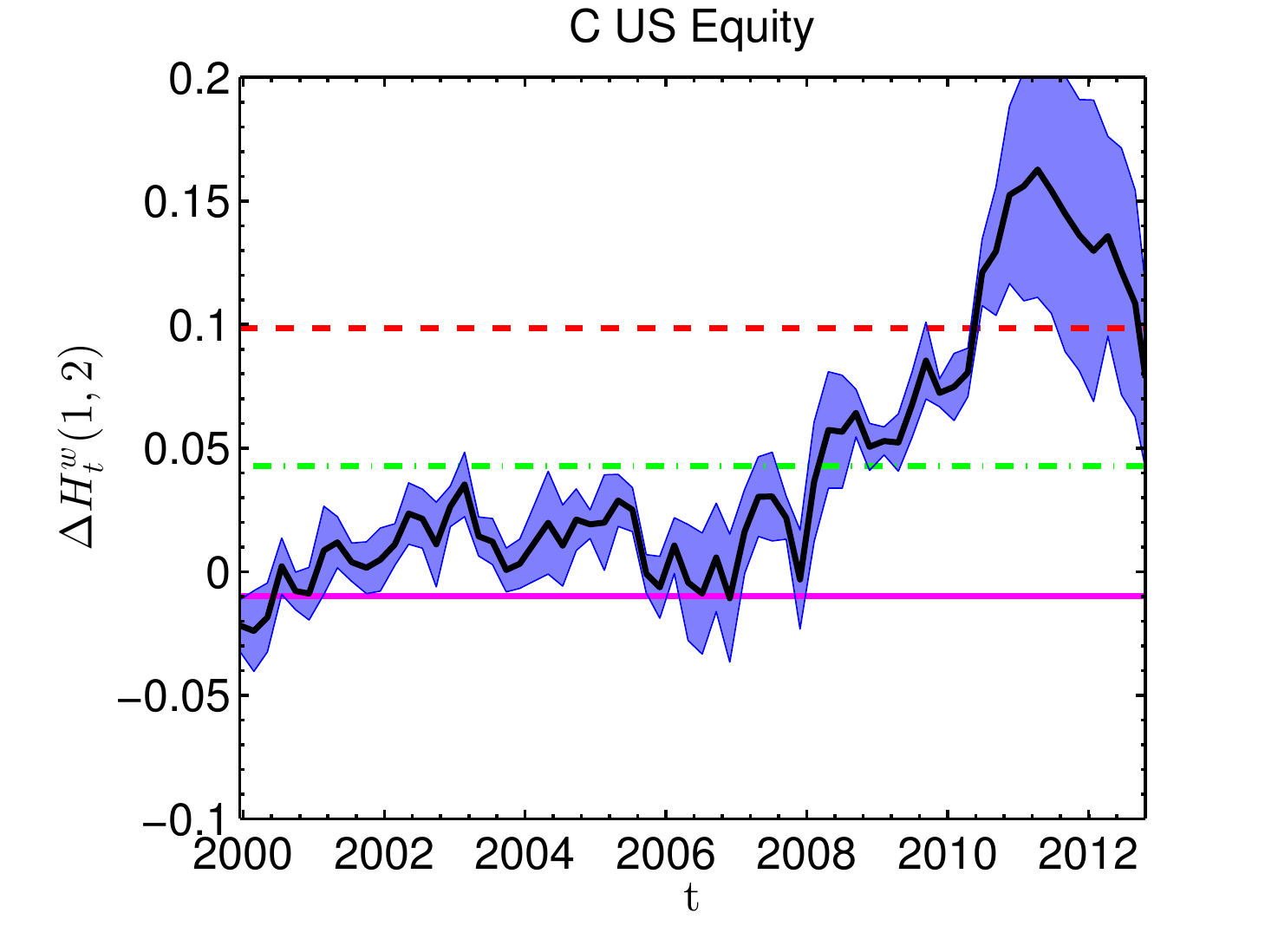} \\
\includegraphics[width= 0.5\columnwidth]{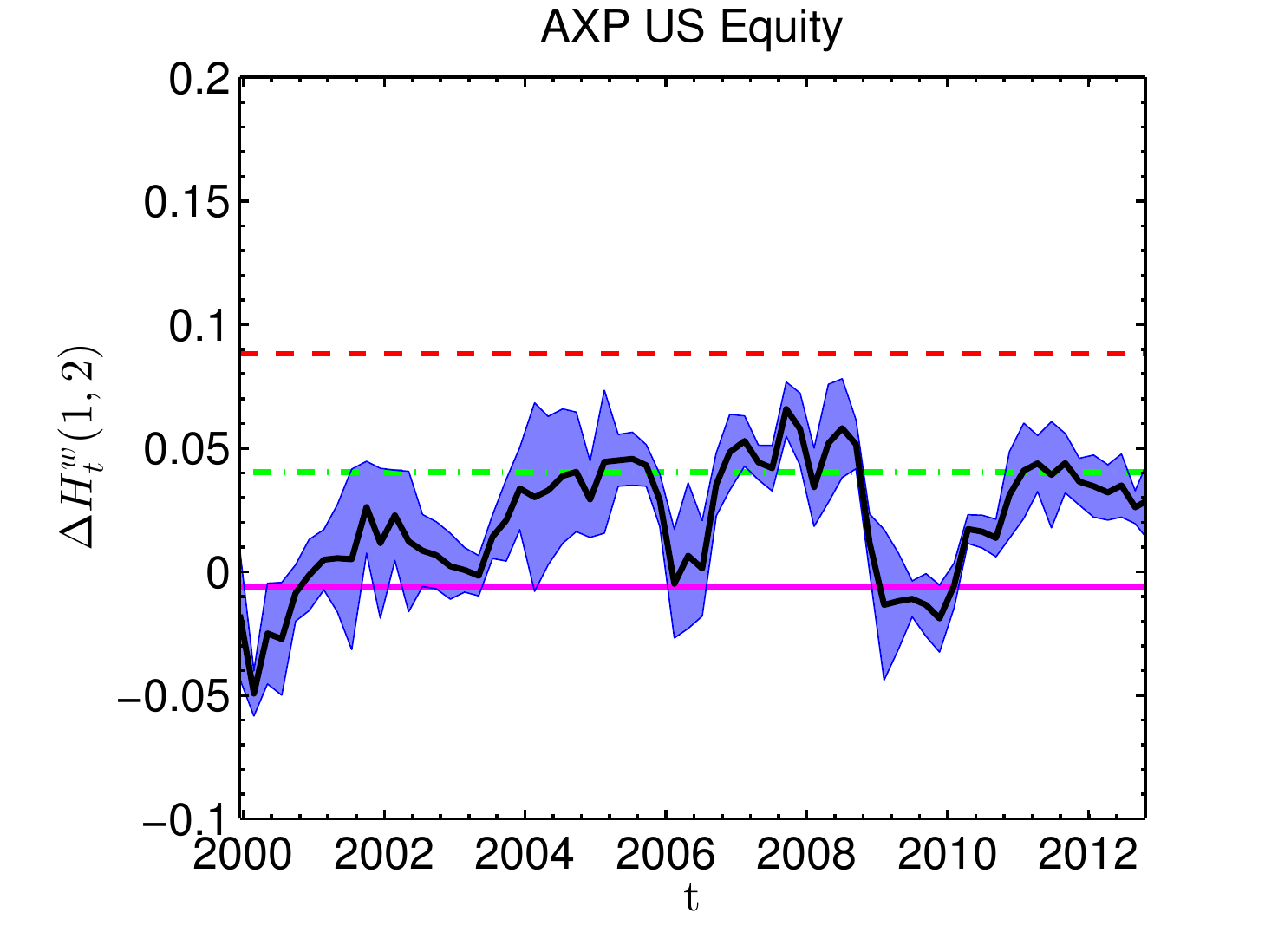} &
\includegraphics[width=0.5 \columnwidth]{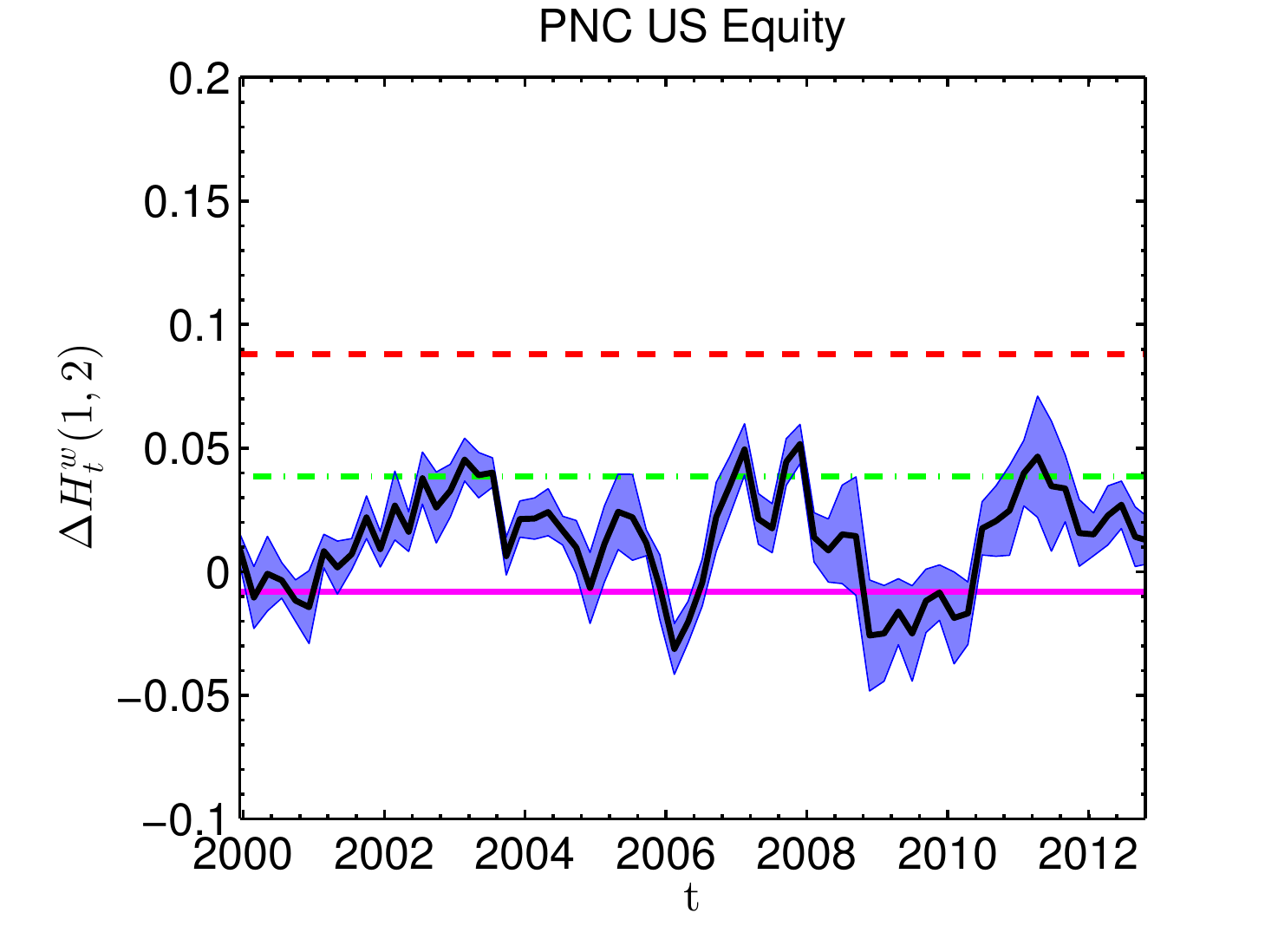}
\end{tabular}
\caption{\label{fig8}
\textit{We plot the dynamical evolution $\Delta H^{w}_{t}(1,2)$ (black thick line) for the six stocks in the time period 01-01-1995 to 22-10-2012. The horizontal lines represent the 2.5\% (red dashed line), 50\% (green dot-dashed line) and 97.5\% (magenta continuous line) quantiles extracted from the distribution of $\Delta H^{w}(1,2)$, obtained from many simulations of MRW with Student t distributed residuals. The blue shaded area represents the error on $\Delta H^{w}_{t}(1,2)$, computed, for each t, as the standard deviation on $\Delta H^{w}(1,2)$. The distributions of the simulated returns have thicker tails than in the original log-normal case, and this results in the quantiles of the sample distributions of $\Delta H^{w}(1,2)$ being larger. } }
\end{figure}

\begin{figure}
\includegraphics[width=\columnwidth]{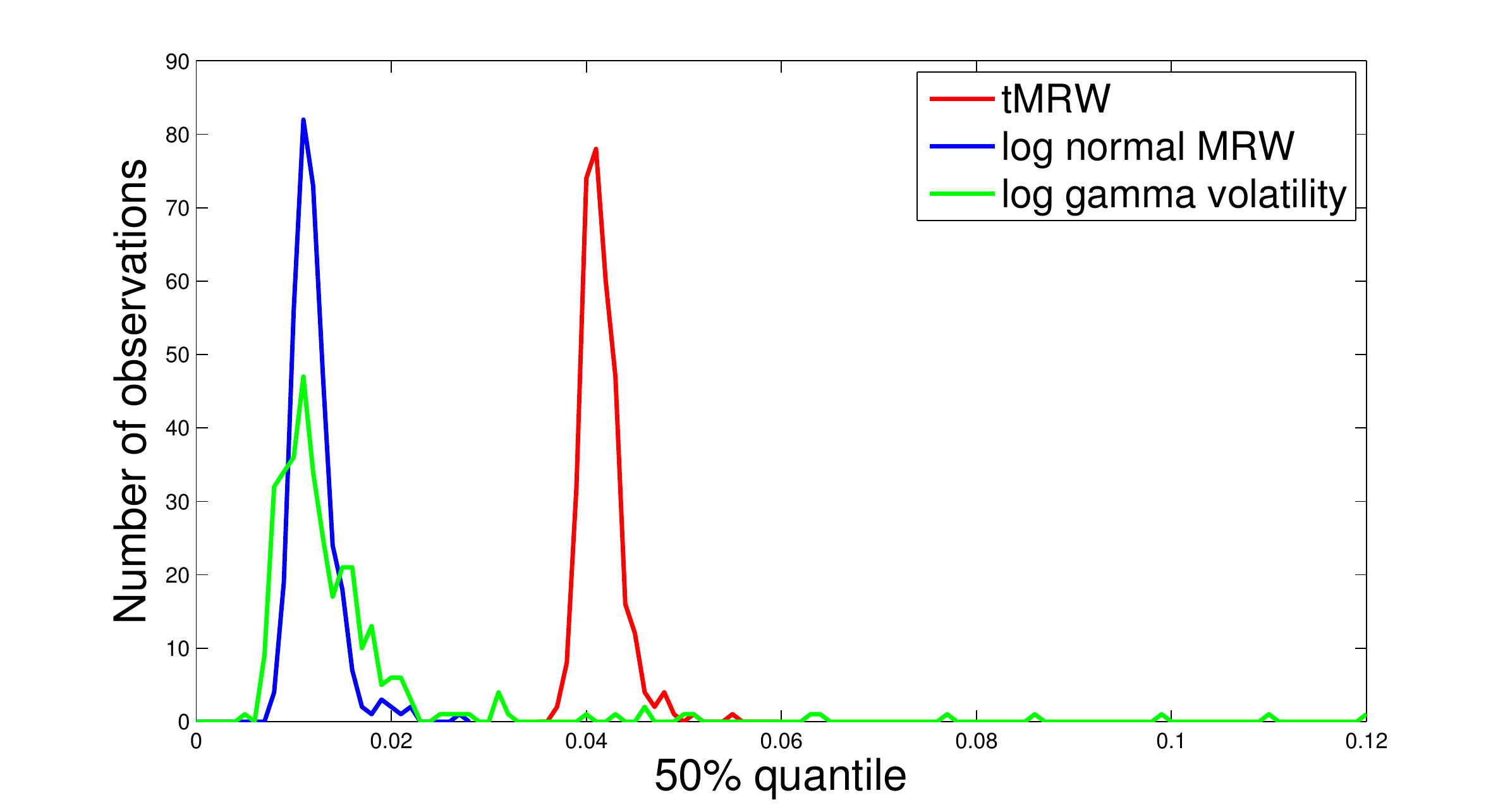} 
\caption{\label{figMedians} \textit{(Color Online) We plot the sample distributions of the $50\%$ quantiles observed on 1000 simulations of the three different specifications of the MRW with the parameters extracted from empirical data: log-normal MRW (blue line), MRW with Student t residuals (red line) and MRW with log-gamma volatility (green line). On the x-axis is the value of the observed quantile, while on the y-axis we report the number of observations. }}
\end{figure}

We have carried out the same analysis performed in the previous section for the complete set of 340 stocks, this time taking the residuals $\epsilon$ to be Student t distributed. MRW time series with Student t residuals (tMRW) have tails fatter than those observed in the log-normal case. We have simulated 1000 realisations of tMRW and computed the proxy of multifractality $\Delta H^{w}(1,2)$ along with the corresponding quantiles. We plot in Figure \ref{fig8} the observed $\Delta H^{w}_{t}(1,2)$ for a selection of stocks together with the quantiles of the distribution of $\Delta H^{w}(1,2)$ computed from the 1000 tMRW simulations. We also show in the top plot of Figure \ref{figHistGamma} the observed percentage of quantile-overpassing for all stocks analysed. As this plot shows, we still retrieve a significant fraction of the observations overpassing the extreme quantiles, with many $\Delta H^{w}(1,2)$ still overpassing the quantiles more than $10\%$ of the time. The rate of quantile-crossing is nonetheless reduced with respect to the case in which residuals are Gaussian. The overall multifractal properties of the simulated time series are very much influenced by the Student t residuals: we observe indeed a systematic shift upwards of the sample distributions of the $\Delta H^{w}(1,2)$'s. As shown in Figure \ref{figMedians}, the median, corresponding to the $50\%$ quantile, is now peaked around $0.04$, whereas the median retrieved from the log-normal case is peaked around $0.015$. 
This tells us that, for a fixed intermittency, the thickness of the tails dramatically rebounds on the multifractal properties of the simulated series, which have now a higher degree of multifractality. This effect, which has also been reported in \cite{barunik2012understanding}, clearly distinguishes different components of the measured multifractal property of the process: on the one hand the non-linear temporal dependence of the series, on the other hand the thickness of the tails of the unconditional distribution. Both aspects contribute to the measured multifractality and, by thickening the tails of the returns distribution, one can account for most of the anomalous fluctuations observed in the scaling exponents. Nonetheless the remaining multifractality still appears to defy the hypothesis of constant volatility covariance. In other words, the presence of more extreme fluctuations in the return process cannot fully account for the anomalous fluctuations observed in the empirical $\Delta H^{w}_{t}(1,2)$. 
\subsection{The effect of log-gamma volatility}\label{secGamm}
\begin{figure}
\begin{tabular}{cc}
\includegraphics[width= 0.5\columnwidth]{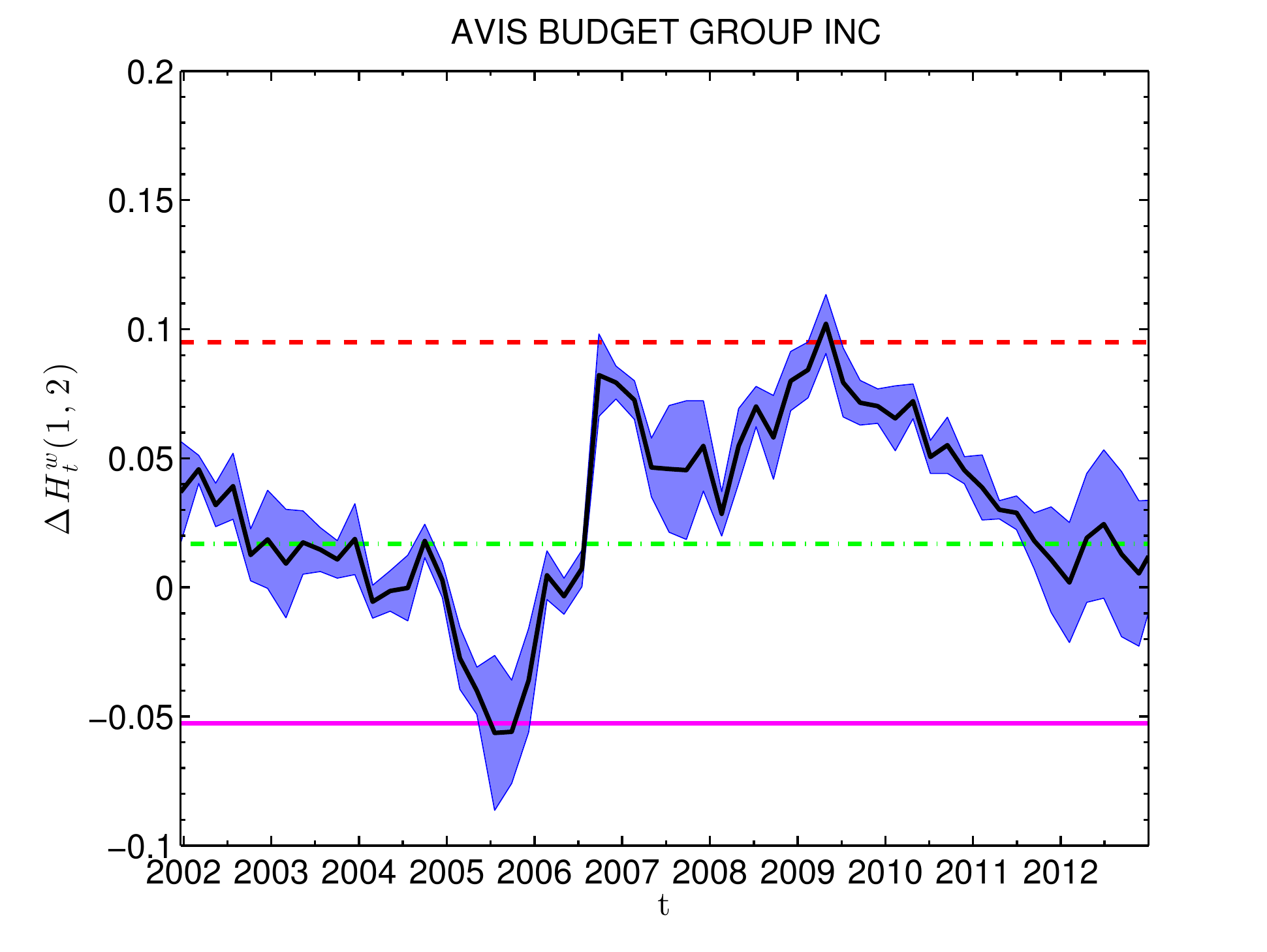} &
\includegraphics[width= 0.5\columnwidth]{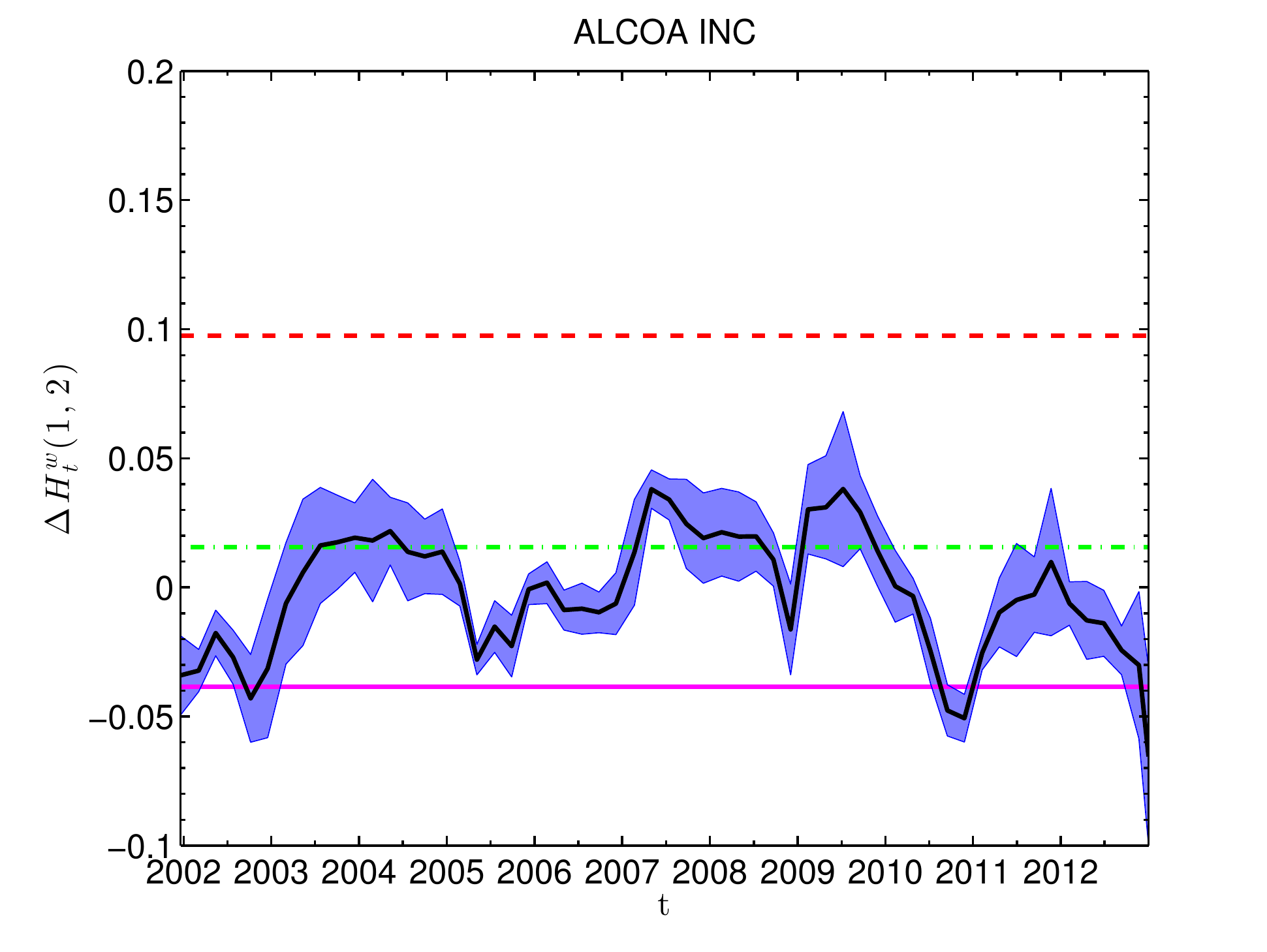} \\
\includegraphics[width= 0.5\columnwidth]{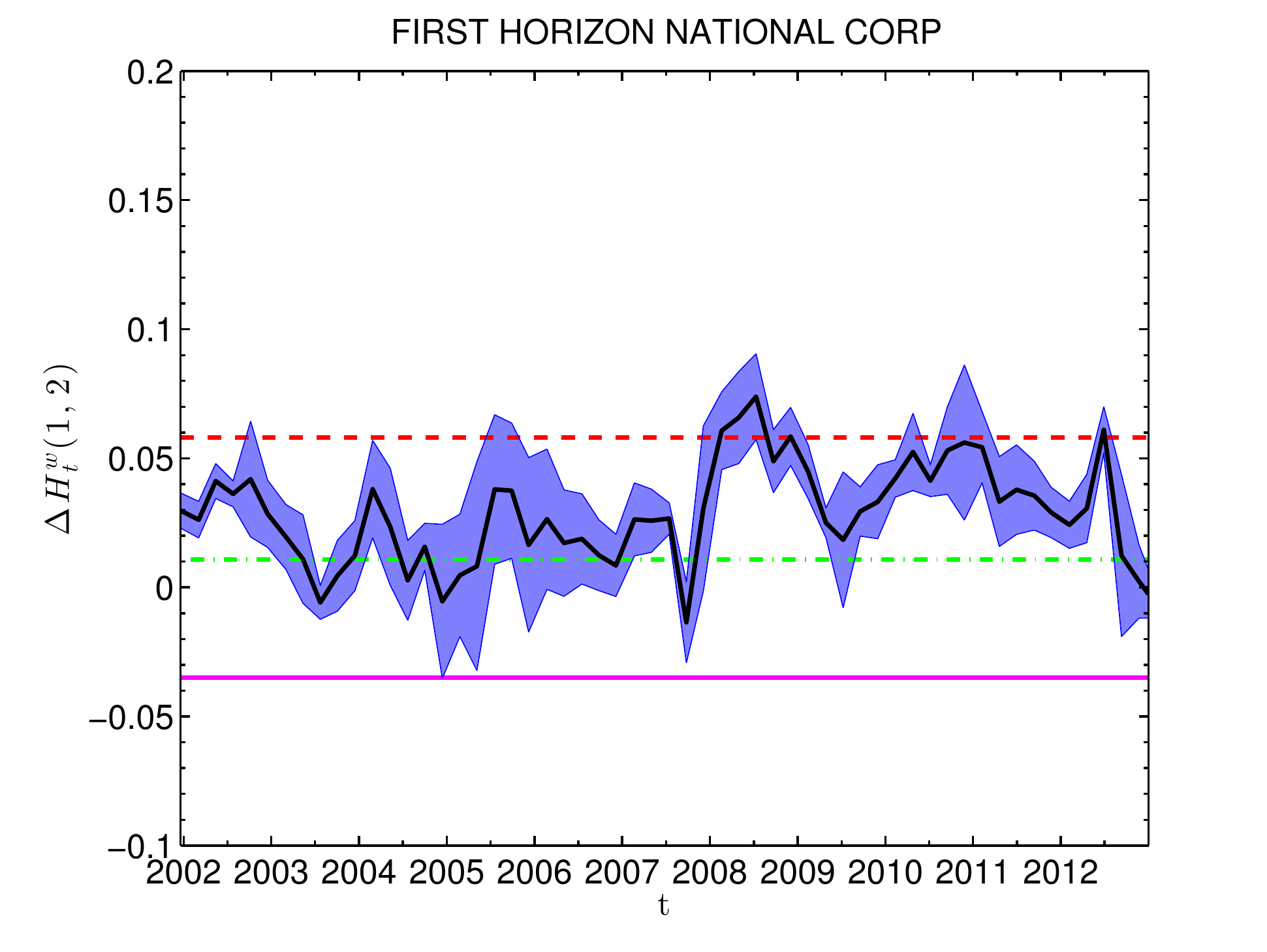} &
\includegraphics[width=0.5 \columnwidth]{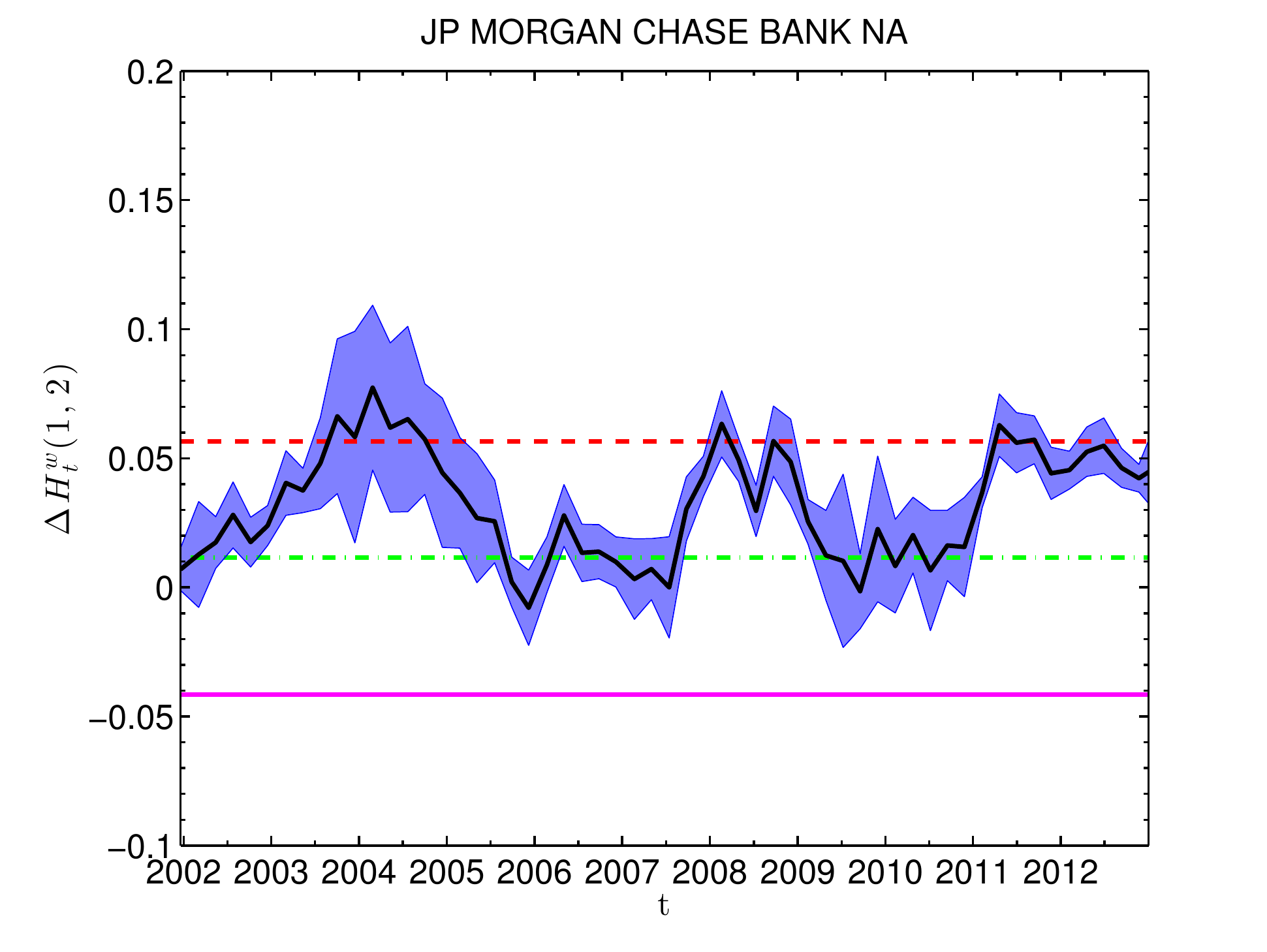} \\
\includegraphics[width= 0.5\columnwidth]{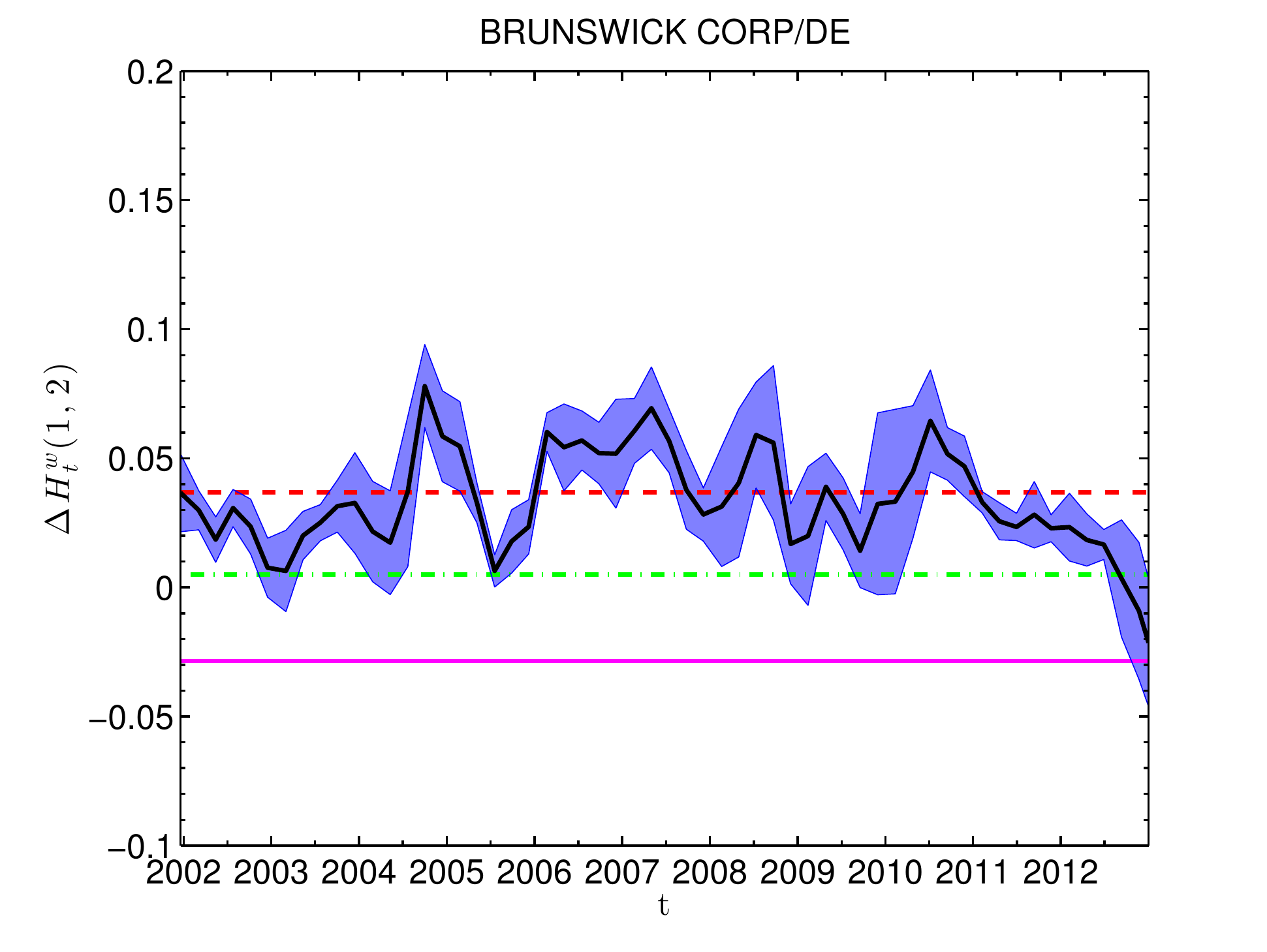} &
\includegraphics[width=0.5 \columnwidth]{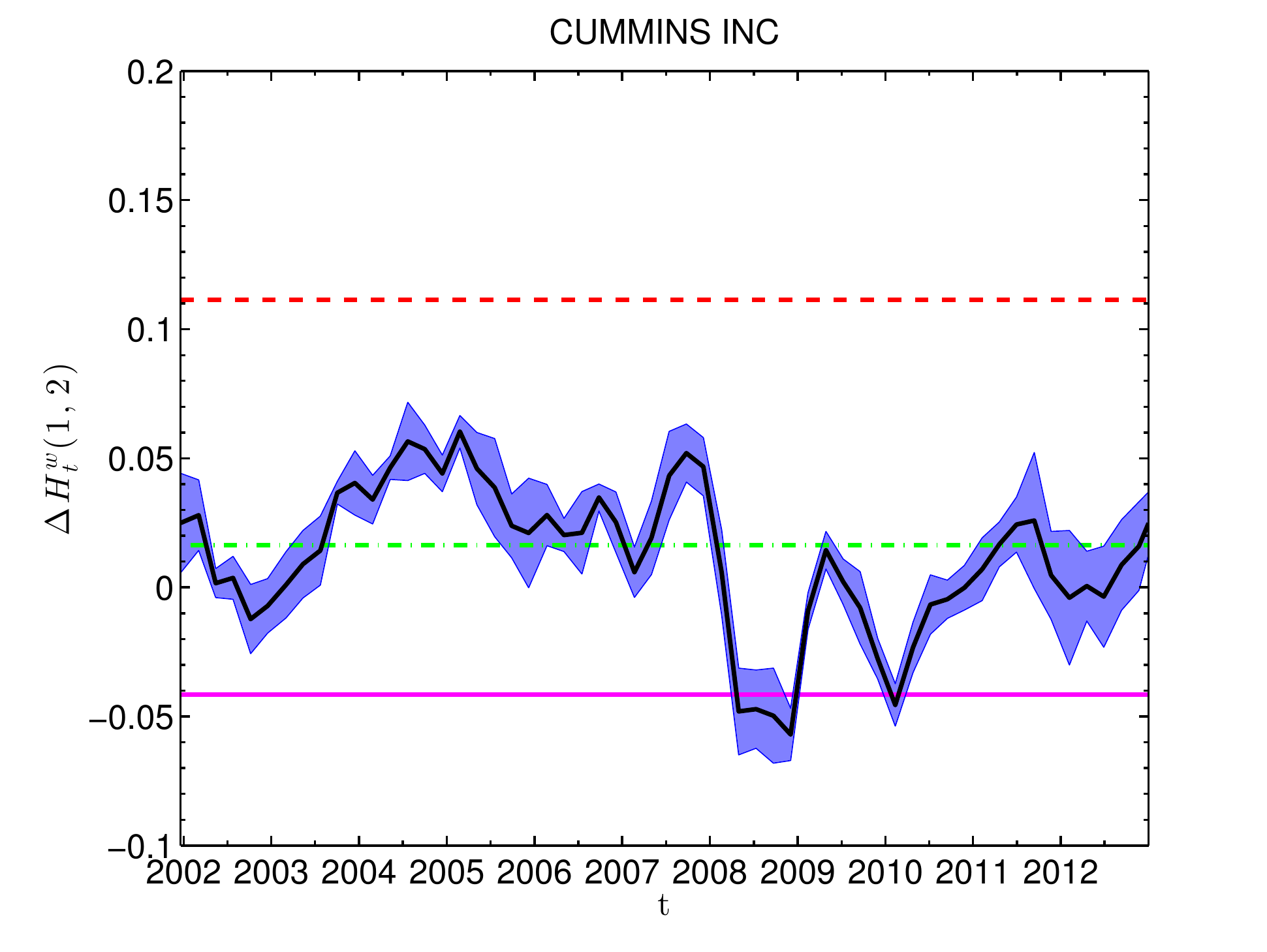}
\end{tabular}
\caption{\label{figGamma}
\textit{We plot the dynamical evolution $\Delta H^{w}_{t}(1,2)$ (black thick line) for six stocks in the time period 01-01-1995 to 22-10-2012. The horizontal lines represent the 2.5\% (red dashed line), 50\% (green dot-dashed line) and 97.5\% (magenta continuous line) quantiles extracted from the distribution of $\Delta H^{w}(1,2)$, obtained from many simulations of MRW with the volatility process $\omega$ simulated as a gamma random variable with $k=\theta=1$. The blue shaded area represents the error on $\Delta H^{w}_{t}(1,2)$, computed, for each t, as the standard deviation on $\Delta H^{w}(1,2)$.} }
\end{figure}

\begin{figure}
\begin{tabular}{c}
\includegraphics[width=\columnwidth]{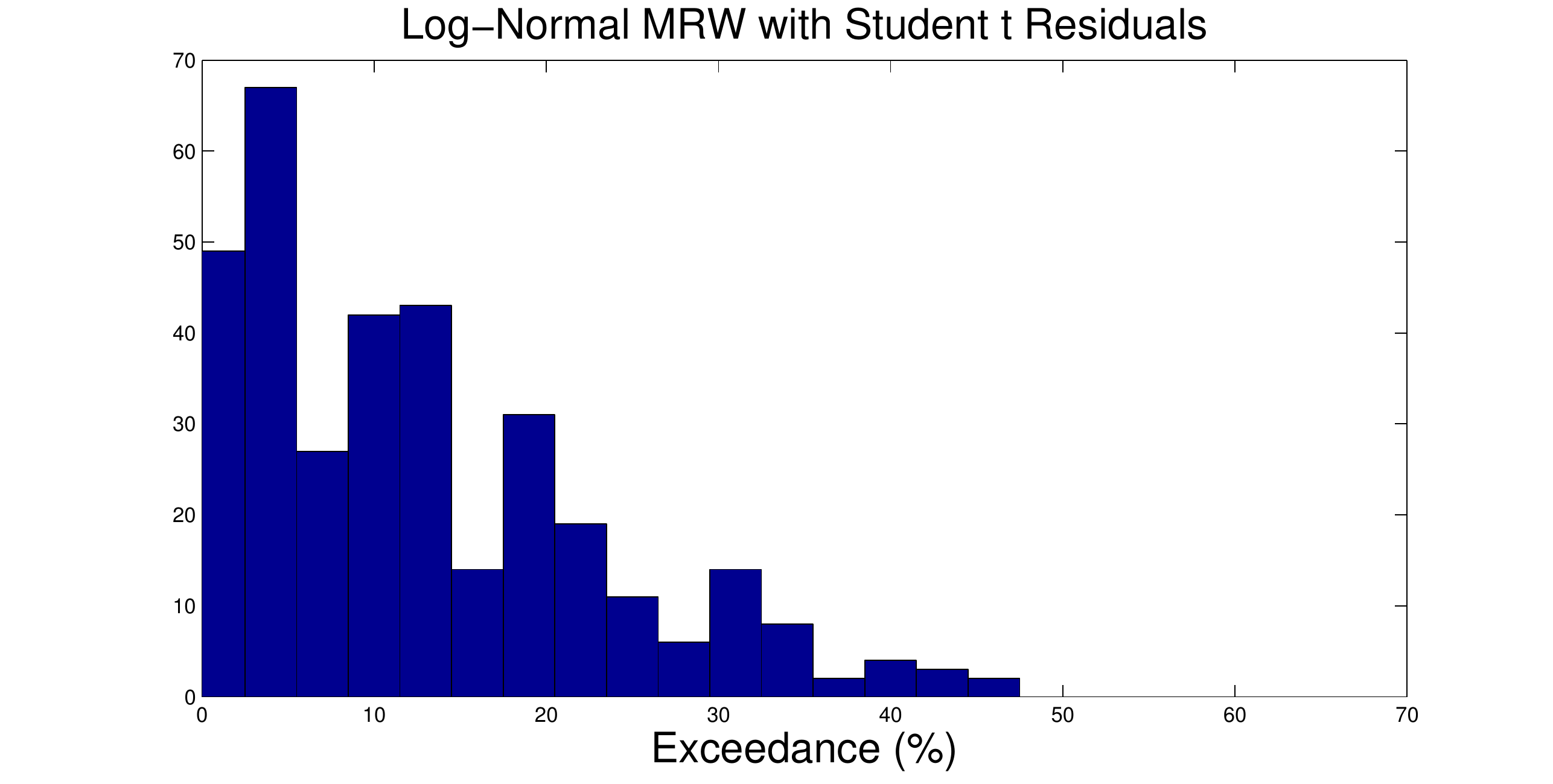} \\
\includegraphics[width=\columnwidth]{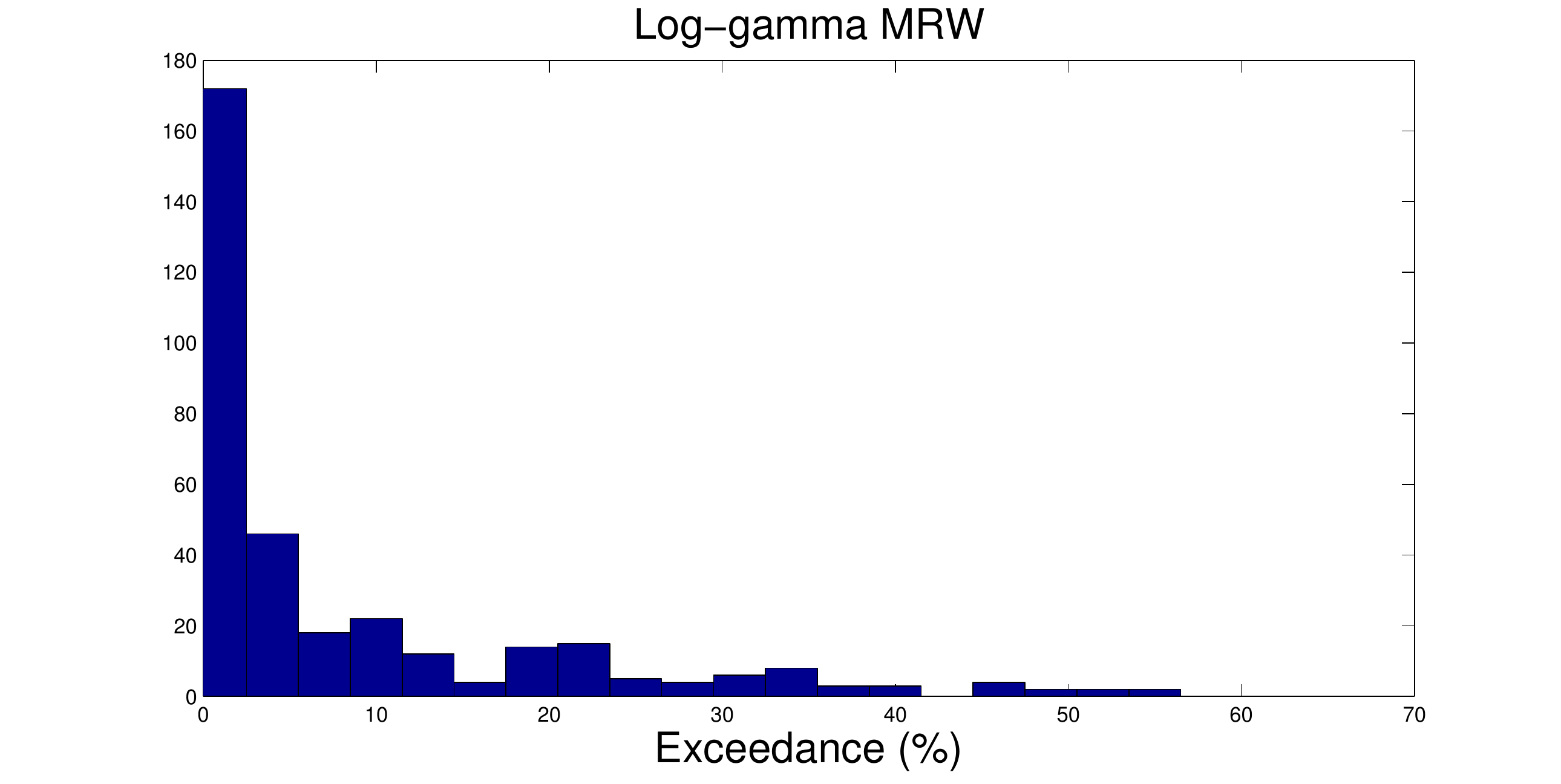}
\end{tabular}
\caption{\label{figHistGamma}\textit{Histogram of the observed exceedance percentages of the whole set of stocks when the benchmark model is taken to be  the log-normal MRW with Student t residuals (top) and the log-gamma MRW (bottom). On the y-axes is the number of stocks exhibiting the percentage of quantile-crossings given in the x-axis. The reduction of the overall number of exceedances is remarkable if compared with that obtained for log-normal MRW, whose histogram is shown in Figure \ref{figHistograms}. }}
\end{figure}

\begin{figure}
\centering
\includegraphics[width= \columnwidth]{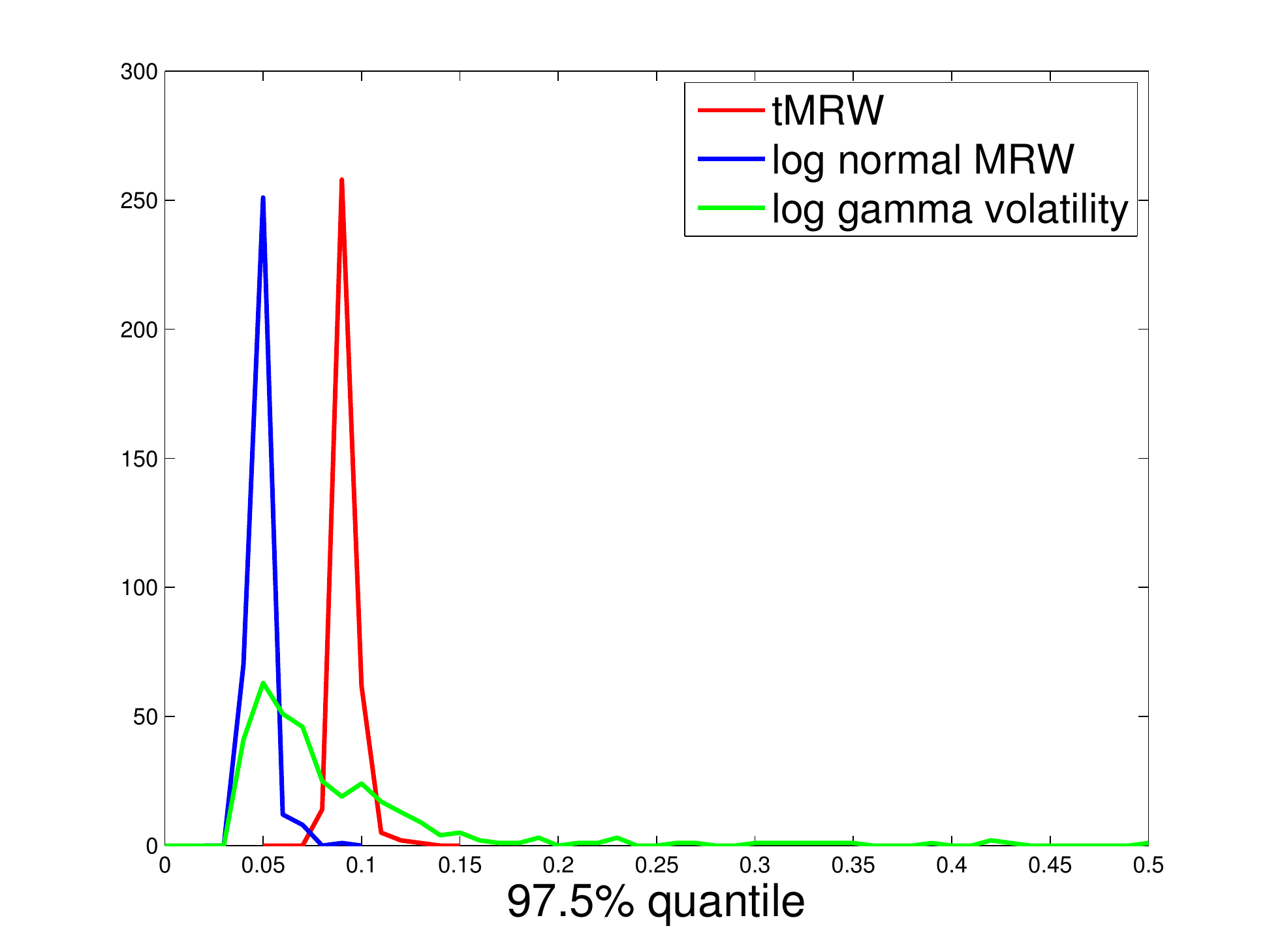}
\caption{\label{fig97Quant}\textit{We plot the sample distributions of the $97.5\%$ quantiles extracted from 1000 simulations of the three different specifications of the MRW with the parameters extracted from each empirical time series: log-normal MRW (blue line), MRW with Student t residuals (red line) and MRW with log-gamma volatility (green line). On the x-axis is the value of the observed quantile, while on the y-axis we report the number of observations.} }
\end{figure}

Another possible modification of the standard log-normal MRW in order to account for fatter tails is to consider the volatility to be log-gamma distributed. It has been shown in \cite{muzy2006extreme} that such a specification can reproduce the fat tails observed in the empirical distribution of stock returns better than the log-normal MRW. Synthetic time series with log-gamma volatility components are obtained replacing the normal distribution with the law given in (\ref{loggamma}) for $\omega$. We have repeated the same testing procedure: after having estimated the model parameters we have computed the $\Delta H^{w}(1,2)$ confidence intervals and, for each stock, we have compared the fluctuations of $\Delta H_{t}^{w}(1,2)$ with the corresponding confidence interval. Examples of $\Delta H_{t}^{w}(1,2)$ for a set of stocks compared with quantiles obtained with this specification of the volatility are shown in Figure \ref{figGamma}. We also show in the histogram in bottom of Figure \ref{figHistGamma} the percentage of quantile-crossing observed on the complete data set. The histogram shows a dramatic reduction of quantile exceedances, with most of the anomalous fluctuations now falling below $2\%$ and therefore in agreement with what envisaged by the log-gamma MRW. As shown in Figure \ref{figMedians}, the median of the $\Delta H^{w}(1,2)$ distribution, although showing some large values, is distributed around the same values observed for the log-normal MRW. What is remarkably more broadly distributed than in the two previous cases is the $97.5\%$ quantile, whose observed distribution on the whole data set is shown in Figure \ref{fig97Quant}, compared with those retrieved for log-normal MRW and tMRW. This confirms that the log-gamma volatility provides a framework in which very large fluctuations of the measured multifractality are much more likely than in the other two cases inspected but, as Figure \ref{figHistGamma} shows, some stocks still show fluctuations of $\Delta H^{w}(1,2)$ which not even the log-gamma MRW can explain.   
\bigskip
\newline Overall we can say that both modifications of the log-normal MRW model confirm that beefing up the tails of the distribution of synthetic time series has a sizeable impact on multifractal properties, with the percentage of anomalous fluctuations being drastically reduced. Nonetheless, we still observe some stocks systematically overpassing the confidence intervals that do not allow to accept globally the hypothesis of constant multifractality.  
\section{\label{sec6}Conclusive discussion}
Main scope of this paper has been to measure and validate variations of multifractality on a set of stock returns. As a benchmark we have considered the MRW model, a parsimonious multifractal model that is able to reproduce faithfully many features commonly observed in stock returns. Since the dynamical estimations of multifractality are made on relatively small samples, to validate the observed fluctuations in the degree of multifractality as truly significant, we have shown that the fluctuations observed in empirical data are truly crossing the extreme quantiles expected from a MRW with constant intermittency coefficient. Confidence intervals have been computed by estimating a proxy of multifractality on different realisations of MRW's simulated with parameters obtained from each empirical time series. We can conclude that the tails of the unconditional distribution of stock returns contribute significantly to the observed fluctuations of multifractality, but cannot fully explain the totality of the quantile crossings. Thus the disagreement between model and data could be possibly explained assuming the intermittency parameter to be time varying, which corresponds to having a breakdown in the serial dependence of the volatility. A possible time varying nature of the intermittency would also affect the tails of the distribution, which depend critically on $\lambda$. We have also shown that increasing the intermittency coefficient makes, as indeed expected, the tails of the GHE distribution computed on the synthetic series thicker, while we found no relevant dependence with $T$. Our analysis thus suggests that a varying intermittency coefficient may be the correct guess towards the inclusion of the observed empirical facts into future multifractal modelling. 
\newline A natural extension, which we are investigating at the present, is the possibility of identifying switching points between multifractality regimes, as suggested by research contributions in regime switching for dynamical correlation \cite{pelletier2006regime,hamilton1990analysis}. As with cross-correlation, one could model the temporal correlation of the volatility to be switching between different regimes of stationarity, although the association of the switching points with realistic economic triggers is yet to be established.  
\newline Starting from these results, our main future aim is the development of a realistic model for stock dynamics which incorporate these empirical findings, possibly backed by a plausible economic ground about what triggers multifractality to change over time.     
\medskip
\section*{Acknowledgments}
The authors warmly thank Jean-Philippe Bouchaud for helpful discussions. We also thank Bloomberg for providing the data. TDM acknowledges support by COST ACTION TD1210.


\end{document}